\documentclass[letterpaper,twocolumn,10pt]{article}

\usepackage{usenix-2020-09}  

\usepackage{tikz}
\usepackage{amsmath}

\usepackage{enumitem}

\usepackage[labelfont=bf]{caption}
\usepackage{subcaption}
\usepackage{cleveref}

\usepackage{amssymb,amsfonts}

\usepackage{algorithm}
\usepackage{algpseudocode}

\usepackage{booktabs}

\usepackage[normalem]{ulem}  

\usepackage[compact]{titlesec}
\titlespacing*{\section}{0pt}{6pt plus 4pt minus 2pt}{2pt plus 2pt minus 2pt}
\titlespacing*{\subsection}{0pt}{4pt plus 2pt minus 1pt}{2pt plus 1pt minus 1pt}
\titlespacing*{\subsubsection}{0pt}{4pt plus 2pt minus 1pt}{2pt plus 1pt minus 1pt}

\usepackage{xspace}
\newcommand{\myparagraph}[1]{
\noindent\textbf{#1.\xspace}}
\newcommand{\myparagraphnodot}[1]{
\noindent\textbf{#1\xspace}}

\newcommand{\eg}{\emph{e.g.}\xspace}

\newcommand{\ie}{\emph{i.e.}\xspace}

\newcommand*{\rom}[1]{\uppercase\expandafter{\romannumeral #1\relax}}

\crefname{insight}{Insight}{Insight}
\newcounter{insight}
\newcommand{\boxinsight}[1]{%
  \refstepcounter{insight}%
  \vspace{\smallskipamount}%
  \noindent \simplebox{
    \textbf{\uline{Insight~\theinsight:}} 
    \textit{#1}
  }
}
\crefname{takeaway}{Takeaway}{Takeaway}
\newcounter{takeaway}
\newcommand{\boxtakeaway}[1]{%
  \refstepcounter{takeaway}%
  \vspace{\smallskipamount}%
  \noindent \simplebox{
    \textbf{\uline{\textit{Takeaway~\thetakeaway:}}}
    \textit{#1}
  }
}

\usepackage[most]{tcolorbox}
\tcbset{textmarker/.style={%
        enhanced,
        parbox=false,boxrule=0mm,boxsep=0mm,arc=3.5mm,
        outer arc=3.5mm,left=2mm,right=2mm,top=4pt,bottom=3pt,
        toptitle=1mm,bottomtitle=1mm,oversize}}

\newtcolorbox{simplenoteBox}{colback=white, colframe=black, boxrule=0.2mm, arc=1.5mm, auto outer arc, boxsep=0mm, left=2mm, right=2mm, top=1mm, bottom=1mm} 
\newtcolorbox{noteBox}{textmarker,
    colback=gray!8!white}

\newcommand{\simplebox}[1]{\begin{simplenoteBox} #1 \end{simplenoteBox}}

\usepackage{ifthen}
\newboolean{publicversion}
\setboolean{publicversion}{false}

\ifthenelse{\boolean{publicversion}}{
    \newcommand{\grumbler}[3]{}
    \newcommand{\esha}[1]{}
    \newcommand{\inigo}[1]{}
    \newcommand{\haoran}[1]{}
    \newcommand{\jx}[1]{}
    \newcommand{\todo}[1]{}
}
{
    \newcommand{\grumbler}[3]{\xspace\textcolor{#3}{\bf #1: #2}}
    
    \newcommand{\haoran}[1]{\grumbler{Haoran}{#1}{blue}}
    \newcommand{\jx}[1]{\grumbler{Jiarong}{#1}{red}}

    \newcommand{\todo}[1]{\textcolor{blue}{TODO: #1}}
}

\newcommand{\sysname}{\textsc{OpScale}\xspace}

\usepackage{titling}

\begin{document}

\date{\vspace{-20pt}}

\title{\vspace{-0pt}\Large \bf \sysname{}: Operator-level Provisioning and Autoscaling for LLM Serving\vspace{-0pt}}

\author{
{\rm Xingqi Cui}\\
Rice University
\and
{\rm Chieh-Jan Mike Liang}\\
Microsoft Research
\and
{\rm Ziang Tang}\\
Rice University
\and
{\rm Jiarong Xing}\\
Rice University
\and
{\rm Haoran Qiu}\\
Microsoft Azure Research
} 

\maketitle

\begin{abstract}
Achieving cost efficiency while meeting strict user-facing SLOs (\eg{}, time-to-first-token) remains a fundamental challenge for cloud GPU clusters serving large language models (LLMs).
Autoscaling is the key mechanism for cluster resource management, yet a basic system design question is open for serving LLMs: \textit{what should be the unit of scaling?}
Existing approaches primarily treat the entire model as a monolithic scaling unit---simple but unable to capture the fine-grained dynamics of inference workloads.
As a result, such coarse-grained scaling often leads to either SLO violations under bursty demand or significant GPU under-utilization.

Our characterization reveals substantial operator heterogeneity, exposing operator-level elasticity as a viable scaling primitive.
We present {\sysname}, a practical \textit{\textbf{operator-level orchestration}} framework of profiling, provisioning, placement, and runtime serving.
{\sysname} is designed to tackle the high complexity and the space explosion problem, arising from operating at this finer granularity.
Evaluated with production traces on up to 40 A100s and 24 GB200s, \sysname{} attains SLOs with up to {36.3\%} fewer GPUs and {28\%} less power, or achieves {44\%} higher throughput under fixed cost budgets.

\end{abstract}

\section{Introduction}

Online serving of large generative models such as LLMs is the most resource-intensive workload in cloud infrastructure today.
Providers must navigate a fundamental tension: meeting strict user-facing Service-level Objectives (SLOs) for Time-to-First-Token (TTFT) and Time-Between-Tokens (TBT), while minimizing GPU footprints and power consumption~\cite{stojkovic2025dynamollm}.

\noindent
\textbf{Motivation.}
To avoid SLO violations, providers often statically provision for peak demand.
However, based on experience from global cloud providers, LLM inference workloads are highly dynamic with hard-to-predict input/output sequence lengths; thus, reserved GPU pools can only be utilized at {50\% and 39\%} when provisioning for the P95 demand of text and multimodal workloads, respectively~\cite{stojkovic2025dynamollm,qiu2025modserve,xiang2025aegaeon}.

\begin{figure}[!t]
    \centering
    \includegraphics[width=\linewidth]{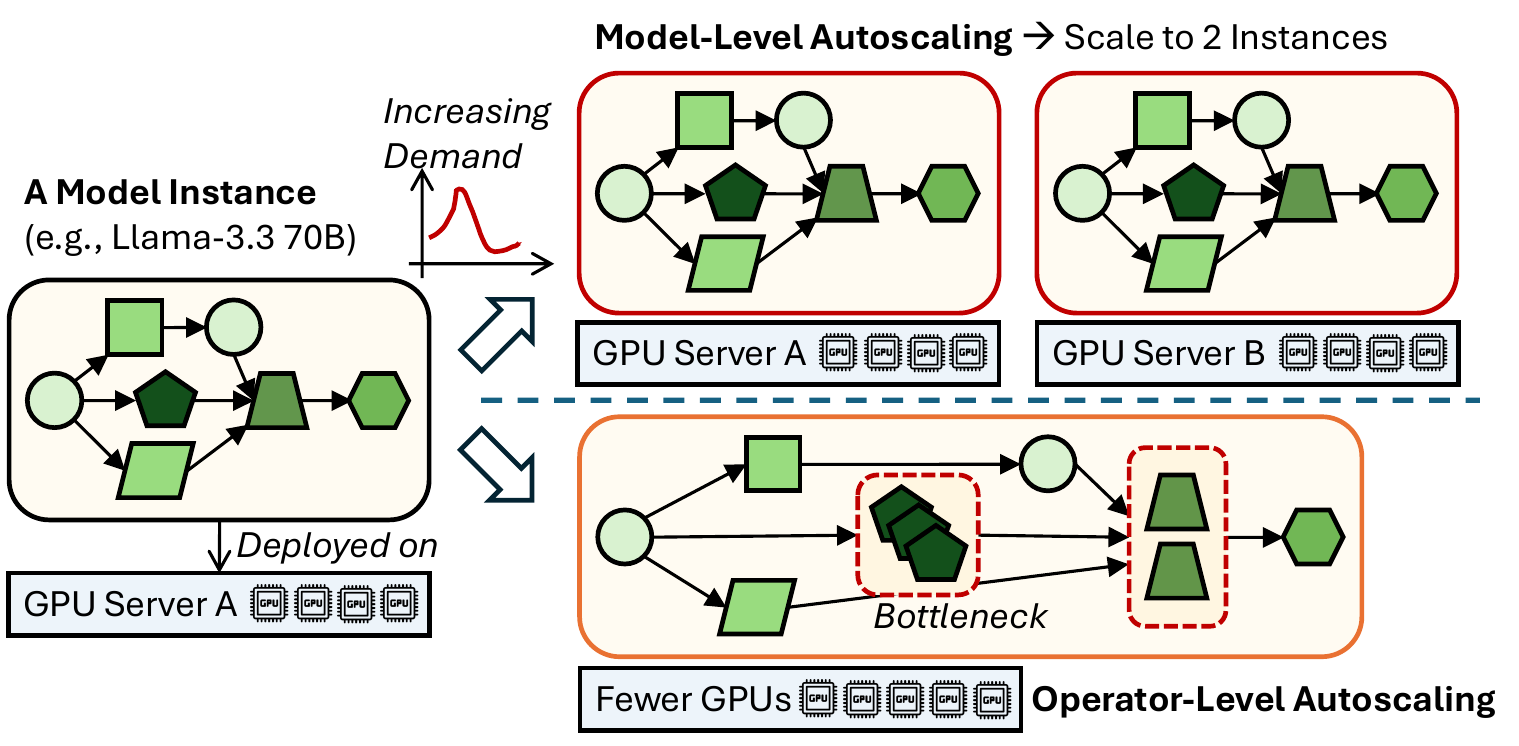}
    \caption{LLM serving at model- vs. {operator-level.}
    }
    \label{fig:idea}
\end{figure}

Achieving high utilization requires SLO-aware \textit{autoscaling}~\cite{stojkovic2025dynamollm,qiu2025modserve,sageserve}.
Today’s state-of-the-art operates at \textit{model-level granularity} (\Cref{fig:idea}, top), scaling the number of full model replicas (\eg{}, prefill instances).
This coarse-grained approach is fundamentally inefficient for two reasons.

\uline{First}, model-level scaling is resource-inefficient.
Generative models are data flows of heterogeneous operators (\eg{}, attention, linear transformation, and normalization).
Forcing all operators to scale uniformly causes \textit{over-provisioning}, where non-bottleneck operators consume precious GPU memory and cycles.
\uline{Second}, model-level autoscaling is slow.
Loading a full model onto additional GPUs incurs significant startup latency (\eg{}, average startup time for a 70B model is around ten seconds even with state-of-the-art methods~\cite{fu2024serverlessllm}).
This delay makes it largely exceed the time granularity of traffic fluctuations, which often occur within seconds in LLM workloads~\cite{qiu2025modserve,stojkovic2025dynamollm}.
Therefore, it either results in SLO violations or costly over-allocation and thus poor cluster utilization.

\noindent
\textbf{Our Work.}
We propose a new paradigm: \textbf{\textit{operator-level}} provisioning and scaling (\Cref{fig:idea}, bottom), which is the key to unblock the two fundamental efficiency limitations.
By shifting the unit of scaling from monolithic models to individual operators within each model replica, we enable precise resource allocation and reduce elasticity latency down to sub-second scales.
However, realizing this paradigm to achieve resource management at such fine granularity requires addressing two systems challenges.

\uline{First}, identifying which operators to scale is non-trivial: 
(1) The inference bottleneck shifts dynamically with workload changes (\eg{}, arrival rate, input lengths). For example, linear operators dominate compute at short sequence lengths, whereas attention operators become the primary bottleneck under long-context workloads;
(2) Not all operators benefit equally from the same resource scaling action. For instance, linear operators gain from added replicas as the arrival rate grows while operators like normalization are lightweight in compute, and benefit more from batching.
\uline{Second}, scaling operators to separate GPUs wastes resources and incurs cross-device communication, whereas colocating scaled operators improves utilization but risks interference across shared GPU resources like SM (Streaming Multiprocessor), memory, and interconnects.
This necessitates interconnect-aware operator placement with fine-grained contention modeling.

To solve these challenges, we first perform a theoretical analysis of operator-level provisioning and scaling to characterize heterogeneity based on workload-aware sensitivity, resource elasticity, and inter-operator communication.
Building upon the theoretical framework, we design \sysname{}, an end-to-end cluster-scale orchestration system across multiple planes.
(1) A \textit{\textbf{data plane}} captures the performance characteristics and resource sensitivity of each operator under diverse workload conditions (batch size, sequence length, query rate);
(2) A \textit{\textbf{control plane}} integrates these profiles with live traffic to compute optimal scaling and contention-aware placement plans;
(3) An \textit{\textbf{execution plane}} deploys provisioning plans, manages distributed operator lifecycles across GPU servers, and orchestrates runtime request routing.
Together, \sysname{} provisions GPU resources at the operator level and scales minimally (\ie{}, the bottleneck operators at traffic spike and the most non-sensitive operators when traffic goes down) to meet fluctuating traffic demands within the running engine.

\noindent
\textbf{Results.}
We evaluated \sysname{} on a cluster of up to 40 NVIDIA A100 GPUs and 24 GB200s, comparing against the model-level inference scaling platforms: DynamoLLM~\cite{stojkovic2025dynamollm}, AIBrix~\cite{team2025aibrix}, and the vLLM Production Stack~\cite{production-stack}.
Across diverse model architectures (\ie{}, text vs. vision models, dense vs. MoE models) on production-scale traces~\cite{stojkovic2025dynamollm,qin2024mooncake} (processed 929K requests and 1.5B tokens), \sysname{} achieves better SLO attainment with up to {36.3\%} fewer GPUs and {28\%} lower power consumption, with gains amplified on a GB200 cluster with faster interconnects (NVL domains)~\cite{gb200azure}.

\begin{figure}[!t]
    \centering
    \includegraphics[width=\linewidth]{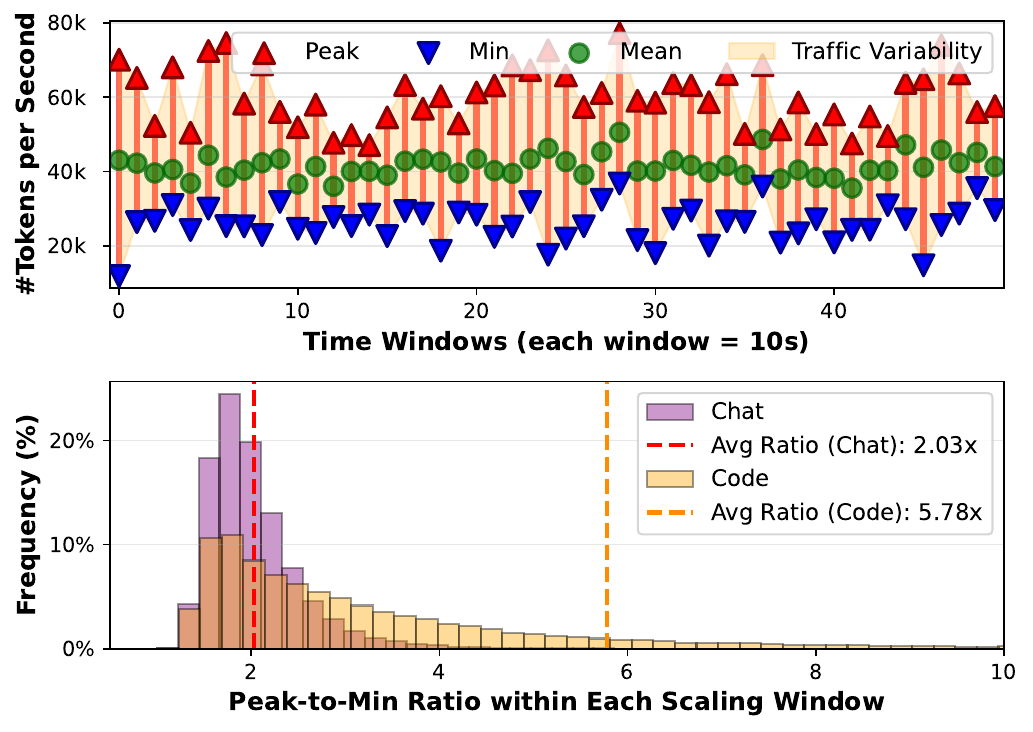}
    \caption{Burstiness in production LLM inference traffic~\cite{stojkovic2025dynamollm} leads to large variability even within a 10-second timescale (for both Chat and Code services), making model-level autoscaling hard to adapt promptly.}
    \label{fig:motivation}
\end{figure}

\noindent
\textbf{Contributions.}
Our main contributions include:
\begin{itemize}[leftmargin=*,nosep]
    \item A systematic characterization of operator heterogeneity on compute–memory sensitivities to workload changes.
    \item A novel and practical operator-level provisioning strategy, to exploit intra-model heterogeneity for high efficiency.
    \item An end-to-end system implementation called \sysname{} that instantiates operator-level resource management.
    \item Comprehensive evaluations with production-scale traces to highlight \sysname{}'s superior GPU cost savings.
\end{itemize}
\begin{figure}[!t]
    \centering
    \includegraphics[width=\linewidth]{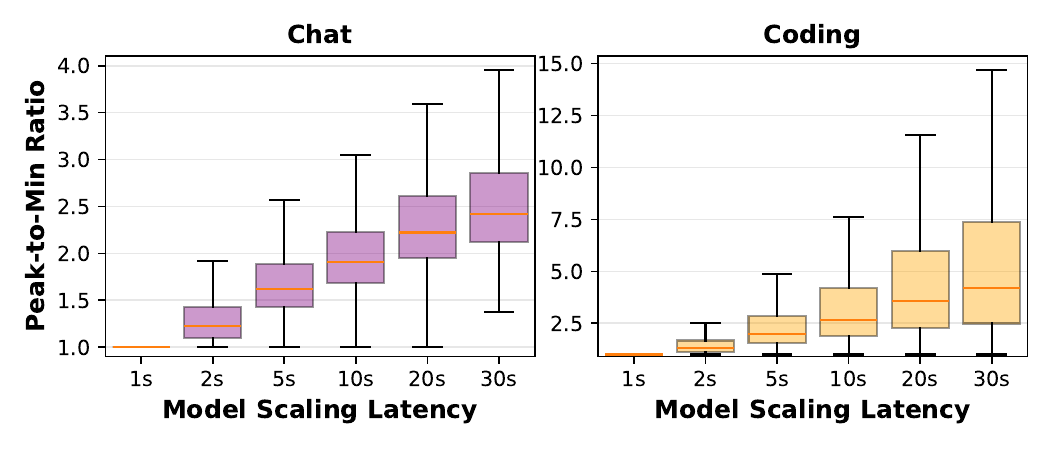}
    \caption{Larger autoscaling latency misses more fine-grained traffic variations across production workloads due to the increased load peak-to-min ratio.}
    \label{fig:scaling-overhead-trend}
\end{figure}

\begin{figure*}
    \centering
    \includegraphics[width=\textwidth]{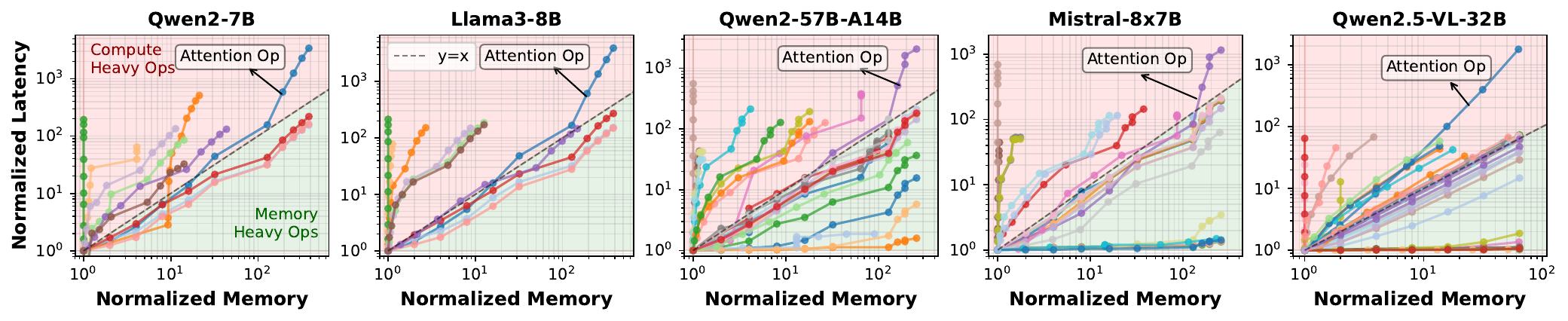}
    \caption{Operator compute and memory sensitivities to sequence lengths across different model architectures. Each line is a different operator, with each data points measured at sequence lengths in [128, 256, ..., 64K] from left to right. The $y=x$ marks the linear scaling with increased sequence length. Detailed per-operator results are provided in \Cref{fig:comp-sensitivity} and \Cref{fig:mem-sensitivity}.
    }
  \label{fig:op-comp-mem-sensitivity}
\end{figure*}

\section{Motivation}
\label{sec:bg}

\sysname{} targets per-model autoscaling, the dominant deployment pattern in production LLM clusters where GPUs are elastically allocated to serve each model based on its load~\cite{stojkovic2025dynamollm,qin2024mooncake,qiu2025modserve}.
A model can serve diverse workloads (\eg{}, chats~\cite{chatgpt}, deep research~\cite{deepresearch}, and vibe coding~\cite{gpt-codex}), each with distinct traffic patterns and strict SLOs on TTFT and TBT~\cite{stojkovic2025dynamollm}.
As a result, the central challenge arises from meeting these performance targets under dynamic workloads while minimizing total resource costs.

Today, the key solution is SLO-targeted autoscaling~\cite{stojkovic2025dynamollm, team2025aibrix, patke2025hierarchical, dynamo, sageserve}.
Its goal is twofold---(1) provisioning minimal model deployment size without violating user SLOs, and (2) doing so quickly to avoid unnecessary over/under-provisioning periods that waste total GPU-hours or power.

\myparagraph{Limitations of Model-level Scaling}
Current state-of-the-art serving systems---including AIBrix~\cite{team2025aibrix}, DynamoLLM~\cite{stojkovic2025dynamollm}, and the vLLM Production Stack~\cite{production-stack}---rely on model-level autoscaling. This approach adjusts capacity by replicating the entire model as a monolithic unit (\Cref{fig:idea}, top). While straightforward to implement, this coarse-grained granularity forces a rigid coupling between model components, fundamentally hindering meeting the two requirements above.

\noindent
\textit{(1) \uline{Speed Mismatch between Model Scaling Overhead and Traffic Dynamics.}}
Loading a full model onto additional GPUs incurs significant startup latency (\eg{}, loading a 70B model takes >10 seconds on average, even with state-of-the-art methods~\cite{fu2024serverlessllm}).
This delay can lag the traffic fluctuations common in LLM workloads~\cite{qiu2025modserve,stojkovic2025dynamollm}, often resulting in SLO violations or over-provisioning.
As shown in \Cref{fig:motivation}, production LLM traffic~\cite{stojkovic2025dynamollm} exhibits rapid, high-amplitude bursts where peak demand can exceed the minimum by more than 2\texttimes{} (for Chat) and 5\texttimes{} (for Code) on average, even within a 10-second scaling window.
\Cref{fig:scaling-overhead-trend} further illustrates how model-level autoscaling can miss even more of these short-lived fluctuations in these workloads, as scaling latency increases.

\begin{figure*}[!t]
    \centering
    \includegraphics[width=\linewidth]{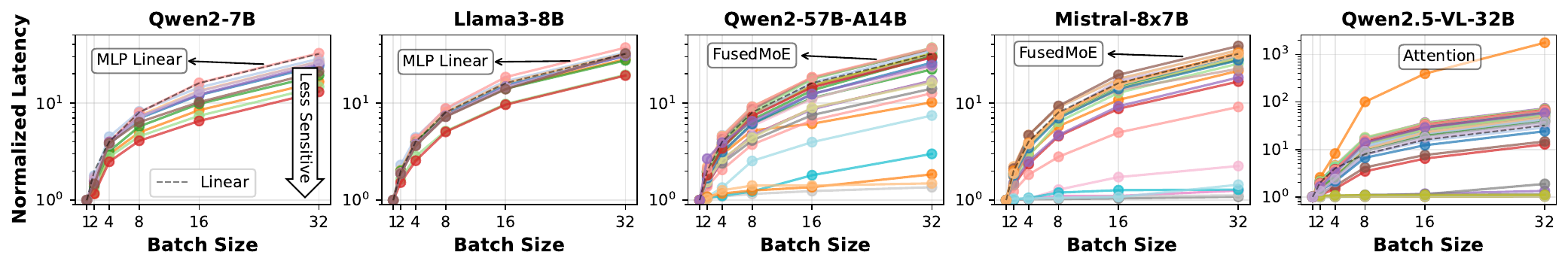}
    \caption{Operator compute sensitivity to batch sizes across different model architectures. Each line is a different operator.}
    \label{fig:batching-sensitivity}
\end{figure*}

\noindent
\textit{(2) \uline{Granularity Mismatch between Monolithic Scaling and Operator Heterogeneity.}}
Model-level autoscaling treats the model as a monolithic scaling unit, ignoring that LLM inference is inherently a data flow graph~\cite{zhu2025nanoflow}, where nodes are operators (\eg{}, attention, linear transformation, and normalization), and edges are data dependencies.
This operator heterogeneity suggests that operators might not equally contribute to SLO violations as traffic demand increases.
Consequently, monolithic scaling can waste resources by uniformly increasing capacity across all operators yet failing to precisely alleviate the truly bottlenecked operators.
To examine this mismatch, \S\ref{sec:bg:characterization} next presents a systematic characterization of operator behavior across diverse model architectures.

\subsection{Operator Characterization Study}
\label{sec:bg:characterization}

To understand how workload dynamics affect different parts of a generative model, we characterize performance metrics for each operator by varying the request sequence length, batch size, and arrival rate. We focus on metrics that can drive autoscaling decisions: compute time and memory usage\footnote{Additional characterization results in Appendix~\ref{sec:appendix:characterization}.}.

Our study characterizes each operator's per-metric \textit{sensitivity}, quantifying the rate at which performance metrics change as workload changes.
We define sensitivity as the normalized performance relative to a base configuration (\eg{}, batch size of one).
It allows us to identify performance-critical operators.

\myparagraph{Experiment Setup}
We consider models of two dominant architectures (\Cref{tab:models}): dense LLMs and mixture-of-experts (MoE). The former includes Qwen2-7B~\cite{qwen2-7b}, Llama3-8B~\cite{llama3}, and Qwen2.5-VL-32B~\cite{qwenvl}. The latter includes Qwen2-MoE~\cite{qwenmoe} and Mixtral-8x7B~\cite{mixtral}.
Models are served by vLLM on a server with NVIDIA A100 GPUs~\cite{a100azure}.
We employ the CUDA time profiler~\cite{nsight} to collect GPU kernel execution times, and instrument inference with a customized version of vLLM's built-in \texttt{layerwise\_profile} library~\cite{vllm-profiler}.
This setup captures fine-grained kernel measurements, including compute time, weight memory, activation memory, and inter-operator communication, enabling a comprehensive analysis of runtime bottlenecks and operator-level summaries.
We use CUDA Green Contexts~\cite{green-context} for the allocation of SM cores to each operator, and DCGM~\cite{dcgm} for SM utilization monitoring.

\myparagraph{Compute Sensitivity}
\label{insight:comp-sensitive}
We profile per-operator compute time as the actual GPU execution time ($\mu$s), recorded using CUDA event profiling. We collect the kernel-only time by excluding CPU overheads and memory-transfer latencies, as these overheads can be effectively hidden by overlapping them with GPU kernel execution using asynchronous operations.

\Cref{fig:op-comp-mem-sensitivity} (Y-axis) and \Cref{fig:batching-sensitivity} present compute-sensitivity curves for all operators in each model, with respect to sequence length and batch size, respectively.
First, we note the non-negligible spread of curves among operators within a model. It highlights how differently the workload changes impact the compute time for each operator.
Second, models exhibit different degrees of heterogeneity, going from a narrower spread (\eg{}, Qwen2-7B and Llama3-8B in \Cref{fig:op-comp-mem-sensitivity}) to a wider spread (\eg{}, Qwen2-57B-A14B and Mixtral-8x7B).
This implies that operators do not equally contribute to SLO violations, especially in the latter case with higher diversity.

The spread of sensitivity curves arises from the intrinsic computational structure of different operators.
As shown in \Cref{fig:batching-sensitivity}, operators with substantial per-token arithmetic (\eg{}, large MLP operations, wide linear projections, and fused MoE operations) exhibit near-linear or even super-linear scaling when batch size increases, as their arithmetic intensity dominates memory and kernel-launch overheads.
In contrast, lightweight operators (\eg{}, layer norms, element-wise activations, small projections) scale sub-linearly because their fixed launch costs and memory-bound behavior become increasingly amortized at larger batch sizes.

Finally, attention operators exhibit the steepest growth: self-attention incurs quadratic complexity in sequence length ($L$), while remaining linear in batch size ($b$). This $\mathcal{O}(L^2 b)$ complexity is significantly higher than operators that scale linearly ($\mathcal{O}(L b)$) with the workload, such as MLPs, norms, and embedding lookups.
Therefore, linear operators (\eg{}, FusedMoE) can dominate compute at short sequences, and attention becomes the primary contributor as sequence length grows.

\boxtakeaway{
\textit{Operators exhibit varying compute sensitivity, with heterogeneity differing across models. 
Workload-aware, operator-level scaling of compute resources is needed rather than uniform, model-wide scaling.}}

\myparagraph{Memory Sensitivity}
\label{insight:mem-sensitive}
We profile both operator weight memory and transient memory usage. The former refers to the static storage of model parameters. The latter refers to the intermediate tensors and KV cache, which are generated during inference computation. By profiling memory usage under varying sequence lengths and batch sizes, we quantify each operator's memory sensitivity based on how its memory usage scales with these factors.
\Cref{fig:op-comp-mem-sensitivity} (X-axis) shows that the attention operator dominates memory growth due to its O($L^2$) scaling with sequence length, while most other operators grow roughly linearly.
Unlike a naïve attention implementation that materializes an $\mathcal{O}(L^2)$ attention-score matrix, the FlashAttention implementation used in our experiments~\cite{dao2022flashattention} avoids it and exhibits approximately linear memory growth with sequence length. Several activation-heavy kernels, such as fused \texttt{act\_and\_mul}, follow a similar trend, whereas operators such as \texttt{index\_select} and \texttt{reshape\_and\_cache} remain nearly flat over the evaluated range of request sequence lengths.

\boxtakeaway{
\textit{Memory sensitivity varies less across operators and models compared to compute sensitivity. It is bounded by linear scaling with FlashAttention.}
}

\begin{figure}[!t]
    \centering
    \includegraphics[width=\linewidth]{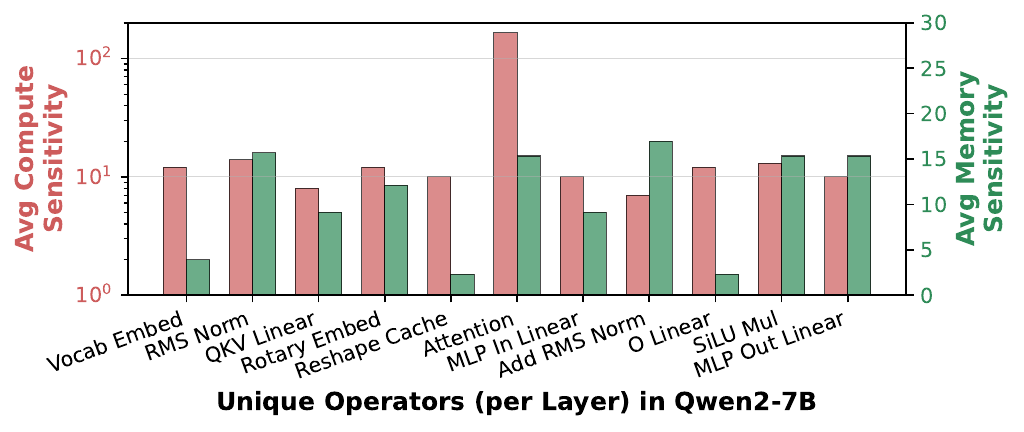}
    \caption{Averaged per-operator compute- vs. memory-sensitive in Qwen2-7B across different sequence lengths.}
    \label{fig:combined-comp-memory}
\end{figure}

\myparagraph{Cross-Dimension Analysis}
\Cref{fig:combined-comp-memory} plots the average per-operator sensitivity, with respect to sequence length for both compute time and memory usage, using a representative layer from the Qwen2-7B model.
We observe substantial variation not only in the absolute magnitude of each sensitivity metric, but also in their relative ratio.
Some operators are highly compute-intensive yet exhibit less memory growth (Attention); others are memory-intensive while remaining lightweight in compute (RMS Norm); and a few are simultaneously heavy or lightweight (QKV Linear) in both dimensions.
This two-dimensional heterogeneity highlights that workload-induced pressure does not manifest uniformly across operators. Instead, each operator responds differently depending on the dominant resource dimension.

\boxtakeaway{
\textit{Operators may be sensitive in both compute and memory, in either, or even in neither.
Due to such mismatch, orchestrating GPU resources requires jointly considering each operator's sensitivity profile and this compute-memory trade-off (\eg{} avoid scaling compute-cheap but memory-heavy operators to save capacity).}
}

\begin{figure}[!t]
  \centering
  \begin{subfigure}[b]{0.48\textwidth}
    \centering
    \includegraphics[width=1\textwidth]{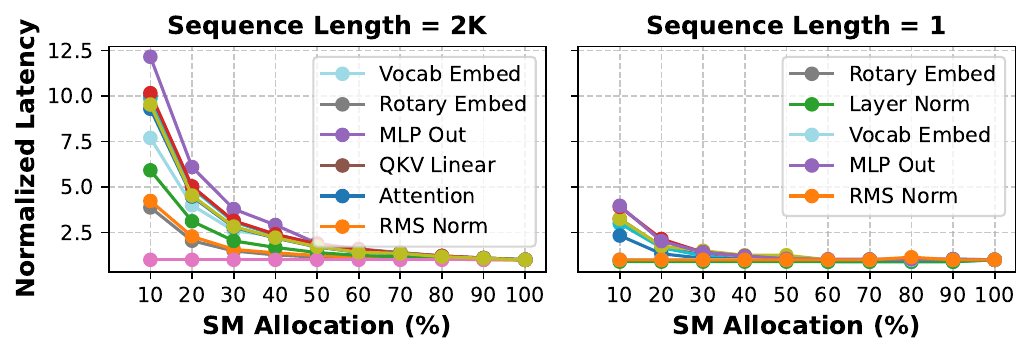}
    \caption{Operator normalized latency to SM allocation.}
    \label{fig:latency-mps}
  \end{subfigure}
  \begin{subfigure}[b]{0.48\textwidth}
    \centering
    \includegraphics[width=\textwidth]{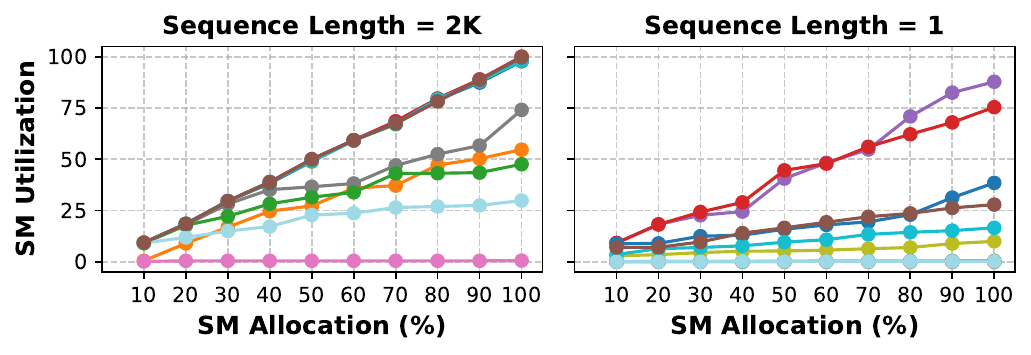}
    \caption{Operator SM utilization (\%) to SM allocation.}
    \label{fig:util-mps}
  \end{subfigure}%
  \caption{Performance and SM utilization characteristics across operators in Qwen2-7B under varying SM allocations.}
  \label{fig:sm-characterization}
\end{figure}

\myparagraph{Sensitivity to SM Core Allocation}
Modern GPUs support spatial sharing, \eg{}, CUDA Green Contexts~\cite{green-context} can specify the number of SMs allocated to each operator's stream.
To guide operator-to-device placement, we characterize the sensitivity of operator compute time to SM core allocation. We consider both the long request sequence of 2K tokens (prefill phase), and the short sequence of 1 token (decode phase).

For the long 2K-token sequence, as SM allocation increases, the normalized latency for all operators decreases significantly (\Cref{fig:latency-mps}). This observation is particularly highlighted for compute-intensive operators such as Attention and MLP. Shown in \Cref{fig:util-mps}, the reason is that these operators naturally require many SM cores. A reduction in SM allocation results in latency increases, hence steeper sensitivity curves.

A different pattern emerges in the decode phase (1-token sequence length). Most sensitivity curves are relatively flat, with a minor latency decrease as SM cores are allocated (\Cref{fig:latency-mps}). Again, \Cref{fig:util-mps} clarifies the reason: the SM utilization is generally too low to saturate the available resources.

\boxtakeaway{
\textit{Sensitivity to SM allocation varies widely across operators, sequence lengths, and prefill/decode phases. Particularly, spatial sharing is suitable for short sequences, as allocation reduction imposes less impact.}
}

\subsection{Rethinking the Scaling Unit}
\label{sec:bg:rethinking}

Rather than improving model-level autoscaling, the intra-layer heterogeneity (\S\ref{sec:bg:characterization}) motivates us to rethink autoscaling at a finer-grained scaling unit---the operators.
Conceptually, the paradigm shift to operator-level autoscaling is feasible, as operators satisfy the desirable property of functional and temporal isolation.
First, operators have explicit execution boundaries during inference. They are compiled into GPU kernels and executed within a CUDA stream, which is an ordered sequence of kernels that may overlap with data transfers to improve throughput~\cite{chen2018tvm}.
These kernels are scheduled to run on the GPU's SM cores. 
Each kernel is run-to-complete without preemption, and there are techniques allowing multiple streams to share SM cores (\eg{}, MPS~\cite{mps}, MIG~\cite{mig}, and CUDA Green Contexts~\cite{green-context}).
Second, each operator maintains its weights and intermediate data in GPU memory, along with temporary buffers for transient memory needs.

\section{Operator-Level Autoscaling}
\label{sec:analysis}

Building upon the insights of operator heterogeneity, a natural question arises: 
\textit{what are the resource efficiency benefits expected from operator (op)-level autoscaling?} To this end, we present our analysis, based on a novel theoretical optimization problem formulation for SLO-driven op-level autoscaling.
Note that solving the optimization problem results in an \emph{offline analytical oracle} to quantify the opportunity from op-level autoscaling.
The online system instead uses the millisecond-scale greedy provisioning heuristic in \S\ref{sec:design:provisioning}, whose resource cost remains within 8\% of this oracle.

\subsection{Theoretical Problem Formulation}
\label{sec:analysis:formulation}

Op-level autoscaling is inherently a multi-objective optimization problem. It considers both SLO and resource objectives, and specifies the following inputs and solutions.

\myparagraph{Problem Inputs and Solutions}
LLM inference is an execution of the operator DAG~\cite{patel2024splitwise}, $\mathcal{G} = (\mathcal{V}, \mathcal{E})$: $\mathcal{V}$ is the set of operators and $\mathcal{E}$ represents data dependencies. 
Given a stream of requests $x\in X$, problem inputs include arrival rate $\lambda$ (QPS) and request sequence length distribution ($L(x)$).
Solutions are per-operator configurations, which include batch size ($B_v$), number of replicas ($R_v$), number of tensor parallelism shards per replica ($P_v$), and per-shard device assignment ($A_v$).
Furthermore, for GPUs that support spatial sharing techniques such as CUDA Green Contexts~\cite{green-context}, the configuration also specifies the number of SMs allocated to a shard ($S_v$).

\myparagraph{SLO Objective}
A request goes through multiple \textit{iterations} of the operator DAG~\cite{patel2024splitwise}. The first iteration is prefill, which processes the full input sequence $L(x)$ of the request. Each subsequent decode iteration generates one output token at a time. Two common inference serving SLOs are TTFT and TBT, for prefill and decode iterations, respectively. The latency of an iteration, $T_{\text{itr}}$, should be less than the SLO.
{\setlength{\abovedisplayskip}{1pt}
\setlength{\belowdisplayskip}{1pt}
\begin{equation}
    T_{\text{itr}} = \sum_{v \in \mathcal{V}} (T_{v} + C_{v} + W_{v}) < T_{SLO}
\end{equation}}\noindent $T_{v}$ and $C_{v}$ are operator $v$'s computation time and communication time to its downstream operators. Both depend on $v$'s configurations ($P_v$, $B_v$, $A_v$) and request sequence length ($L$).
$W_{v}$ denotes the waiting time at each operator $v$ modeled in networked queues~\cite{kiessler1980simulation}, and each operator can be modeled as an $M/M/R_v$ queue~\cite{stability} (Poisson arrivals and exponential service times).
The service rate of $v$ is $\mu_v = \frac{1}{T_{v}}$. Then, $W_v$ is:
{\setlength{\abovedisplayskip}{1pt}
\setlength{\belowdisplayskip}{1pt}
\begin{equation}
    W_{v} = \frac{C(R_v, \rho_v)}{R_v \mu_v - \lambda} \quad \text{with} \quad \rho_v = \frac{\lambda}{R_v \mu_v}
\label{eq:queueing}
\end{equation}
}\noindent where $C(R_v, \rho_v)$ is the Erlang-C formula~\cite{bonald2012internet}.

\myparagraph{Resource Objective}
To minimize the logical shard-replica demand across all operators:
{\setlength{\abovedisplayskip}{1pt}
\setlength{\belowdisplayskip}{1pt}
\begin{equation}
    \min \sum_{v \in \mathcal{V}} P_v \cdot R_v
\end{equation}
}
\noindent
At the same time, deploying operator replicas needs to respect the memory and SM constraint of each GPU device ($M_d^{\text{cap}}$).
{\setlength{\abovedisplayskip}{1pt}
\setlength{\belowdisplayskip}{1pt}
\begin{equation}
\label{eq:memory-capacity}
    \sum_{v \in \mathcal{V}: A_v = d} M_v \;\le\; M_d^{\text{cap}}, \sum_{v \in \mathcal{V}: A_v = d} S_v \;\le\; 100,
    \quad \forall d \in \mathcal{D}
\end{equation}
}\noindent where $M_v = M^{\text{weight}}_v + M^{\text{transient}}_v$, or weight memory and transient memory (\eg{}, activations, KV cache).

\subsection{Benefit Analysis}
\label{sec:analysis:results}

Building on the analytical formulation in \S\ref{sec:analysis:formulation}, we use a profiling-driven performance model that captures each operator’s compute cost, memory footprint, communication volume, and queueing behavior. {$L(x)$ is derived from production LLM inference traces~\cite{stojkovic2025dynamollm}.}
This model\footnote{{Accuracy of operator profiles and the analytical model is in \S\ref{sec:eval:prediction-accuracy}.}} computes the expected TTFT, TBT, and energy consumption for any operator configuration and identifies the optimal SLO-compliant deployments by exhaustively enumerating all feasible configurations.
Our analysis utilizes two representative LLMs: dense Qwen2-7B and MoE Qwen2-57B-A14B. We vary the workload from 10--100 QPS and 128--64K sequence lengths, and evaluate a spectrum of stringent to relaxed SLOs.
We compare the minimal SLO-compliant provisioning plan at the model level with the optimal op-level provisioning plan using the analytical model, solved using exhaustive search.

\begin{figure}[!t]
    \centering
    \includegraphics[width=\linewidth]{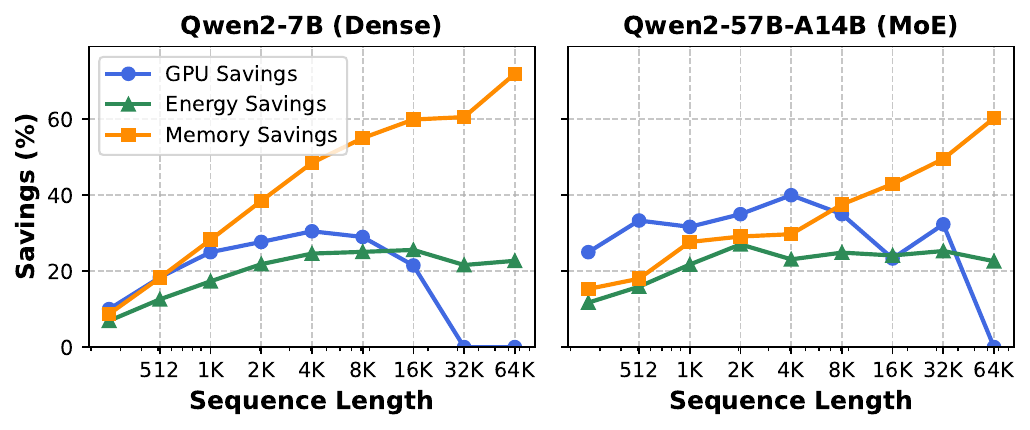}
    \caption{Benefits of op-level provisioning compared to model-level under varying sequence lengths (refer to \Cref{fig:analysis-seqlen} in Appendix for full results).}
    \label{fig:analysis-seqlen-combined}
\end{figure}

\myparagraph{Varying Sequence Length}
\Cref{fig:analysis-seqlen-combined} shows that GPU savings (in blue)  peak around 30\% for dense model and 40\% for MoE. This peak occurs at 4K-token sequences. The savings start to drop at 8K, where the compute load grows to saturate SM cores and limit GPU sharing. MoE models exhibit higher savings because their operators have a more diverse sensitivity.
Energy savings (in green) follow a similar trend, reaching up to 25\% reduction at peak. Unlike GPU device savings, energy savings are still noticeable beyond 4K sequences. The reason is that op-level autoscaling only needs to scale a subset of operators, in contrast to model-level autoscaling.
In addition, GPU memory savings (in orange) grow steadily with sequence length, surpassing 60\% at 32K sequences.

\boxtakeaway{
\textit{Op-level autoscaling is most beneficial at moderate sequence lengths, though long sequences can still exhibit up to 25\% energy and 60\% memory savings.}
}

\begin{figure}[!t]
    \centering
    \includegraphics[width=\linewidth]{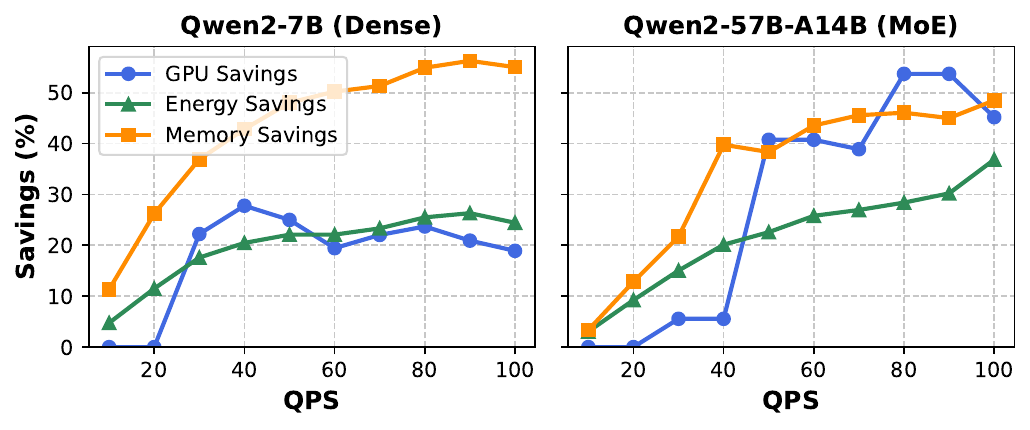}
    \caption{Benefits of op-level provisioning compared to model level under varying QPS (refer to \Cref{fig:analysis-qps} in Appendix for full results).}
    \label{fig:analysis-qps-combined}
\end{figure}

\myparagraph{Varying QPS}
\Cref{fig:analysis-qps-combined} (in blue) shows that GPU device savings peak around 30\% near QPS = 40. 
Beyond this point, savings fluctuate slightly, as SM saturation and tighter latency margins reduce opportunities for GPU device sharing. On the other hand, savings remain negligible at low QPS (<20), where the provisioned model instance can handle limited traffic demands without scaling.
Energy savings (in green) reach up to 25\% at high QPS for Qwen2-7B, primarily driven by the reduced GPU device count from op-level autoscaling's selective operator scaling decisions.
Memory savings (in orange) grow steadily with QPS, surpassing 50\% at 100 QPS, reflecting a similar trend to sequence length growth.

\boxtakeaway{
\textit{Op-level autoscaling yields the most benefits at moderate to high QPS. The finer-grained provisioning mitigates queueing delays, without excessive allocations.}
}

\begin{figure}[!t]
    \centering
    \includegraphics[width=0.95\linewidth]{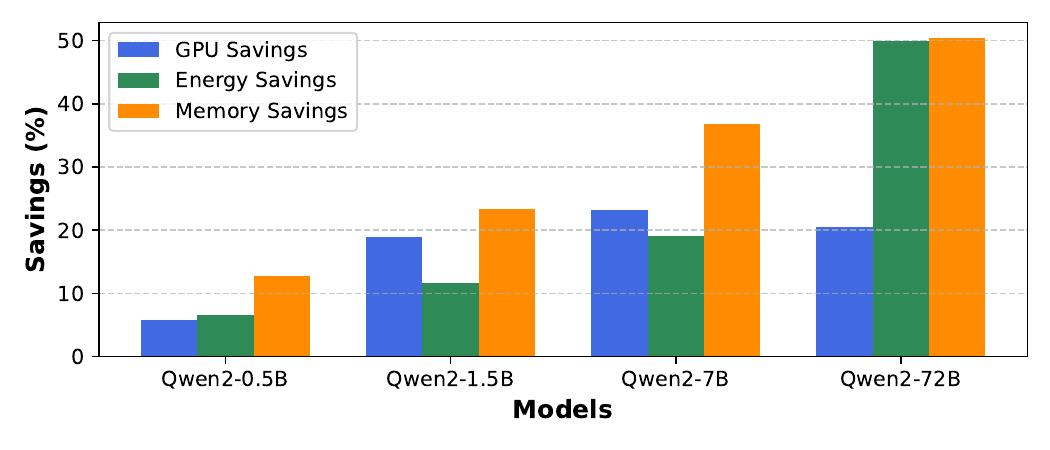}
    \caption{Saving comparisons across model sizes.}
    \label{fig:analysis-layers}
\end{figure}

\myparagraph{Large vs. Small Models}
We analyze the factor of model size using the Qwen2 model family, ranging from 0.5 to 72 billion parameters.
As shown in \Cref{fig:analysis-layers}, op-level scaling yields substantial savings, especially as the model size scales up. For smaller models, savings are generally modest. The reason is that operators do not exhibit a high degree of heterogeneity, which necessitates going finer than model-level scaling.
However, as model size increases to Qwen2-1.5B and Qwen2-7B, GPU and energy savings rise to $\sim$20--30\%, while memory savings exceed 35\%. For the largest model, Qwen2-72B, both energy and memory savings peak at $\sim$50\%. Large models tend to exhibit a higher degree of operator heterogeneity, encouraging the use of finer-grained resource provisioning\footnote{Additional analysis results on op-level provisioning (including a study on prefill and decode stages) are in Appendix~\ref{sec:appendix:prefill_vs_decode}.}.

\boxtakeaway{
\textit{Larger models benefit more from op-level autoscaling, due to greater operator heterogeneity.}
}

\section{\sysname{} Design and Implementation}
\label{sec:design}

\begin{figure}[!t]
    \centering
    \includegraphics[width=\linewidth]{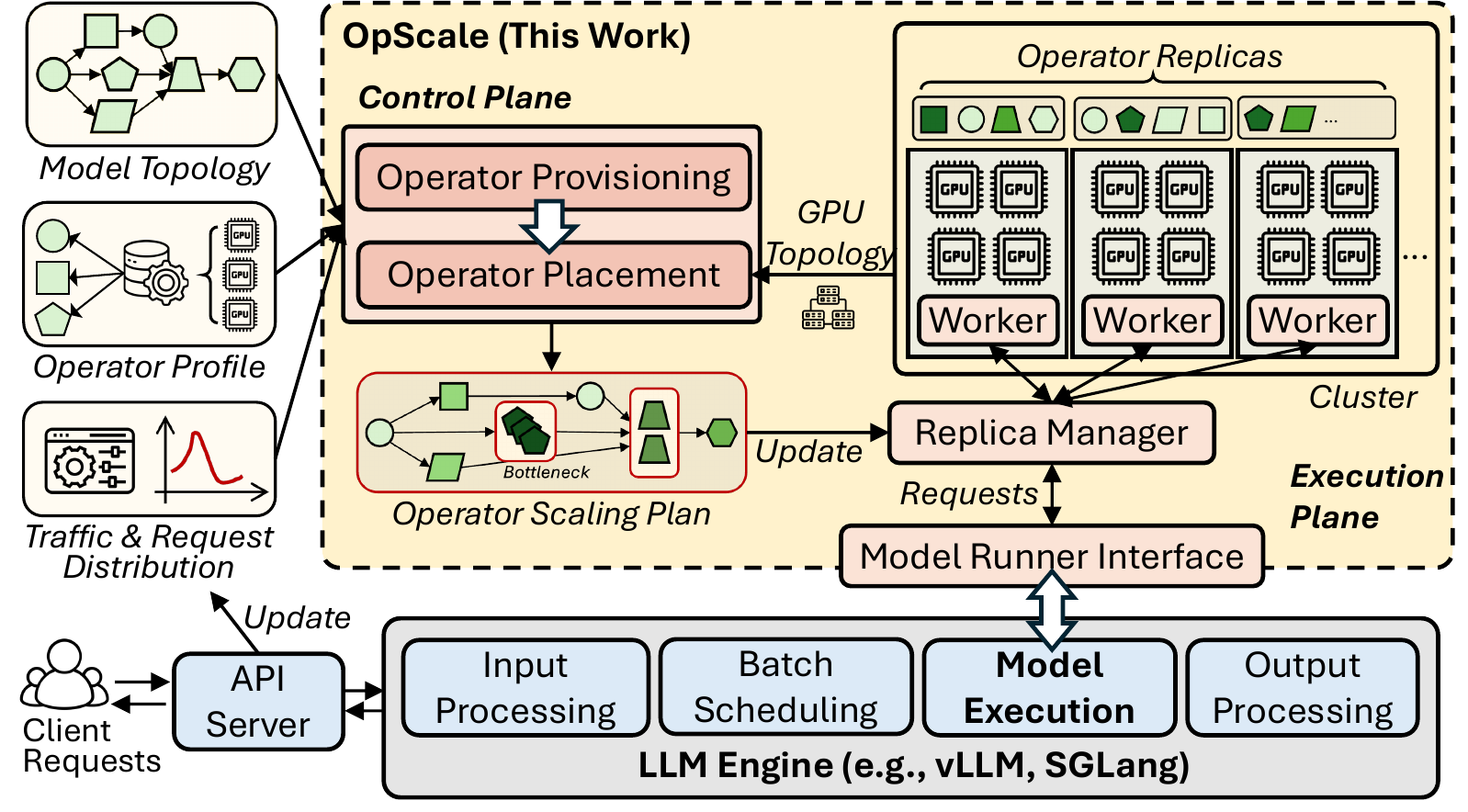}
    \caption{\sysname{} architecture overview.
    }
    \label{fig:overview}
\end{figure}

Realizing op-level scaling in practice requires overcoming significant profiling, optimization, and runtime challenges at such fine granularity. Specifically, the finer granularity results in a growing number of scaling units. To this end, we design a system called \sysname{}, to tame this resulting complexity with two decoupled planes (\Cref{fig:overview}).

\myparagraph{Control Plane}
As the runtime core of {\sysname}, the control plane operates asynchronously to compute an \textit{operator scaling plan} (specified in \S\ref{sec:analysis:formulation}) through three modules:
\begin{enumerate}[nosep,leftmargin=*]
    \item \textbf{Operator Profiler} (\S\ref{sec:design:profiling}): An offline module that profiles three operator attributes: \textit{(1)} computation time ($T_v$),
    \textit{(2)} memory usage ($M_v$), and
    \textit{(3)} communication overhead ($C_v$) from its topological dependencies.
    Profiling is a one-time effort and can be reused across model architectures.
    
    \item \textbf{Operator Provisioning} (\S\ref{sec:design:provisioning}):
    A module that computes optimal operator configurations ($P_v, R_v, B_v$) based on operator profiles, current traffic, and model topology.
    
    \item \textbf{Operator Placement} (\S\ref{sec:design:placement}): A module that maps logical replicas to physical GPUs ($A_v, S_v$), accounting for resource contention and cross-device communication.

\end{enumerate}

\myparagraph{Execution Plane}
A centralized \textit{Replica Manager} (\S\ref{sec:design:runtime}) orchestrates replica lifecycles and serves as a request dispatcher.
Unlike model-level routing, which assigns each request to a full model replica, \sysname{} routes requests through pools of operator replicas: each stage gathers outputs from its replica pool and forwards the activations to the next stage's operators.
{This routing employs a weighted shortest-queue policy, accounting for SM allocation and colocation contention.}

\subsection{Complexity and Tractability}
\label{sec:design:challenges}

The complexity from a growing number of scaling units is reflected in the combinatorial explosion of operator profiling and configuration spaces.
While \S\ref{sec:analysis:formulation} provides the theoretical basis for optimal op-level autoscaling, applying it directly to a runtime system can be impractical, especially with the prohibitive cost of running exhaustive search. We discuss the two main space explosion problems next.

\myparagraph{Operator Profiling Space}
Characterizing operator sensitivities ideally requires profiling operator behavior ($T_v, M_v, C_v$) across a vast state space, which is defined by batch size ($B$), sequence length ($L$), and SM allocation percentage ($S$). 
The problem arises as we consider their ranges. A typical setup has $B \in [1, 256]$, $L \in [1, 65536]$ (discretized), and SM allocation percentage $S \in [1, 100]$. Here, a naive exhaustive search would require profiling over $10^7$ configurations \textit{per operator}. Even if each micro-benchmark run takes only 100~ms, exhaustively profiling a 0.5B-parameter model (with 11 unique operators) would require \textit{weeks} of GPU time. This is untenable for model iteration and deployment in production.

\myparagraph{Operator Configuration Space}
Deriving the optimal operator scaling plan is an NP-hard integer non-linear programming problem (\S\ref{sec:analysis:formulation}).
The search space grows combinatorially with the number of operators and each of the configuration dimensions ($P_v$, $R_v$, $S_v$, $A_v$).
Brute-force solvers take \textit{minutes} for even a small 0.5B-parameter model, while production traffic requires high-quality decisions within seconds (\Cref{fig:motivation}).

\subsection{Operator Profiling}
\label{sec:design:profiling}

Our experience with exhaustive profiling highlights an observation, where it ignores patterns in operator behavior. Considering compute sensitivity, while \Cref{fig:batching-sensitivity} shows that operators can have seemingly different sensitivity curves with respect to batch size changes, they mostly exhibit a monotonic trend.

The observation above suggests that it is feasible for Operator Profiler to employ a \textit{sparse sampling} strategy.
Instead of performing exhaustive enumeration, we first sample random points in the $(B, L)$ space. These points are then used by techniques such as piecewise interpolation. Interpolation is simple, as it is non-parametric and does not require training prior to estimating an unsampled point.

Furthermore, we exploit the \textit{structural equivalence} of LLM architectures. As some operators (\eg{}, Attention and MLP) are identical across layers and models, their profiles can be reused.
With parallel execution of sampling tasks, multi-billion-parameter models can be profiled in under an hour.

\subsection{Operator Provisioning}
\label{sec:design:provisioning}

\begin{figure}[!t]
    \centering
    \includegraphics[width=0.95\linewidth]{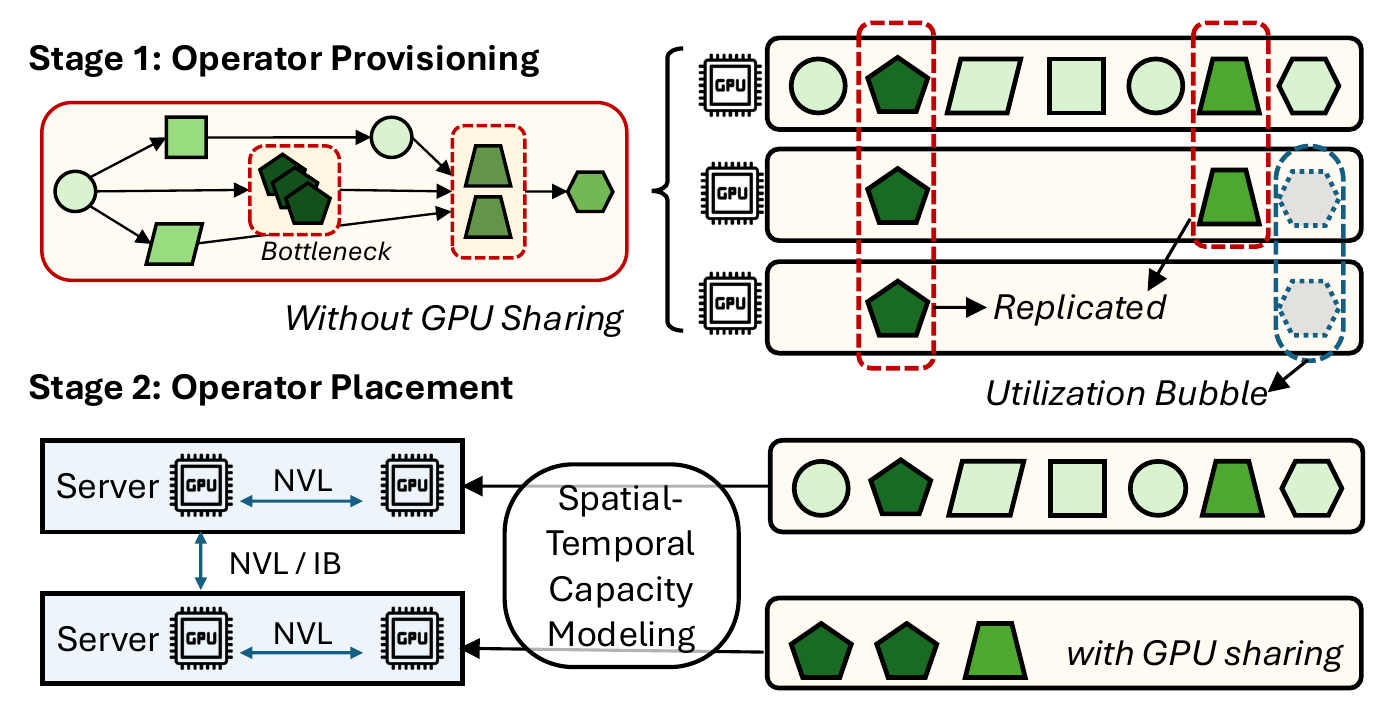}
    \caption{\sysname{} control plane consists of operator provisioning (\S\ref{sec:design:provisioning}) and placement (\S\ref{sec:design:placement}).}
    \label{fig:stage1-stage2}
\end{figure}

To navigate the combinatorial explosion challenge, {\sysname} leverages an observation derived from operator heterogeneity. Specifically, sensitivity curves suggest that operators experience different degrees of performance impact due to workload changes.
In other words, by prioritizing critical operators (\ie{}, operators that expect the most impact) for provisioning first, we avoid exhaustively provisioning \textit{all} operators in the model.

This optimization process proceeds in two distinct phases\footnote{\Cref{alg:greedy-autoscale} in Appendix \ref{sec:appendix:analysis}.}:

\begin{enumerate}[topsep=0pt,parsep=0pt,leftmargin=*]
    \item \textbf{Initialization (Local Optimality):}
    We first establish a baseline configuration by inheriting the initial parallelism $P_v$ from the model deployment strategy (\ie{}, tensor parallelism for intra-server and pipeline parallelism for cross-server deployment~\cite{vllm-tp-pp}). For each operator, we scan feasible batch sizes $b \in \{1,\ldots,B_v^{\max}\}$ to find the minimal replica count $R_v(b) = \lceil \lambda / \mu_v(b,P_v) \rceil$ required to sustain the request arrival rate with no queue build-up (Eq.~\ref{eq:queueing}). We select the $(B_v, R_v)$ pair that minimizes the local sojourn time $T_v+W_v$, providing a latency-optimal starting point.

    \item \textbf{Iterative Critical Path Optimization:}
    The algorithm then evaluates the global DAG latency $T_{\text{total}}$ against the SLO by iteratively identifying the current critical path:
    \begin{itemize}[topsep=0pt,parsep=0pt,leftmargin=*]
        \item \textbf{Upscaling (SLO Violation):} If $T_{\text{total}} > \text{SLO}$, we identify the bottleneck operator $v$ that contributes the most to the critical path latency in the DAG.
        We greedily apply the specific adjustment: either increasing replicas $R_v$ or tuning $(B_v, P_v)$ that yields the maximal latency reduction per unit of resource cost. Repeat until SLO is met.
        \item \textbf{Downscaling (Resource Reclamation):} Conversely, if $T_{\text{total}} < \text{SLO}$, we relax the configuration of the top non-critical operators to reclaim resources, ensuring the system operates at the efficiency Pareto frontier.
    \end{itemize}
\end{enumerate}
Unlike the exhaustive search used only for the offline benefit analysis (\S\ref{sec:analysis:results}), this greedy method runs online and achieves near-optimal resource efficiency: its resource cost is within 8\% of the brute-force oracle, while online incremental plan generation takes only milliseconds (\S\ref{sec:eval:overhead}).

\subsection{Operator Placement}
\label{sec:design:placement}

\sysname{} treats placement as a multi-dimensional bin-packing problem, mapping operator replicas onto minimal GPUs. For GPUs supporting spatial sharing, \sysname{} can further reduce the total GPU footprint, by colocating operators. However, this requires the placement algorithm to be aware of the colocation interference. This section first discusses our interference model, and then presents a locality-aware \textit{Best-Fit Decreasing} heuristic with spatial-temporal modeling\footnote{\Cref{alg:greedy-placement} in Appendix~\ref{sec:appendix:analysis}.}.

{
\myparagraph{Interference Model}
As operators share a device via CUDA Green Contexts, contention on HBM bandwidth, L2 cache, and SM schedulers introduces performance interference.
Let operator replica $v$ run on device $d$ with batch size $b$ and SM fraction $p$; we define an \textit{interference factor}:
{\setlength{\abovedisplayskip}{1pt}
\setlength{\belowdisplayskip}{0pt}
\begin{equation}
\label{eq:interference}
    I_{d, v}(b, p) = \tilde{T}_v(b, p) \,/\, T_v(b, p) \;\ge\; 1,
\end{equation}
}where $\tilde{T}_v(b, p)$ is the co-execution latency and $T_v(b, p)$ is the isolated latency at the same SM allocation.
This formulation captures co-location effects in a single empirical factor $I_{d, v}(b, p)$, which is used by Alg.~\ref{alg:greedy-placement} without additional modeling assumptions.
During the one-time offline profiling effort, we sweep sequence length and SM splits at 5\% granularity for each operator-type pair, co-running both operators on separate green-ctx streams.
The resulting $I_{d,v}$ values are stored in a lookup table; at placement time, queries take $O(1)$ time, with configurations snapped to the nearest profiled grid point.
Our model covers operator \textit{pairs}, approximating three-way co-location by multiplying pairwise factors.
We validate prediction accuracy against held-out measurements in \S\ref{sec:eval:prediction-accuracy}.

\myparagraph{Best-Fit Placement Algorithm}
With $I_{d,v}$ defined, the placement heuristic proceeds in three steps:
}

\begin{enumerate}[nosep,leftmargin=*]
    \item \textbf{Decomposition \& Ordering:}
    To simplify the packing problem, we first decompose the global resource demand into two sets: (1) \textbf{Base Instances} ($\mathcal{R}_{\text{base}}$), the minimum full-model replicas required for baseline service stability, deployed monolithically; and (2) \textbf{Elastic Components} ($\mathcal{R}_{\text{extra}}$), the remaining unassigned operator replicas to be packed dynamically. We then sort $\mathcal{R}_{\text{extra}}$ in descending order of computational cost $T_v$ (or dominant resource demand). This prioritizes bottleneck operators when the cluster has the most contiguous space, reducing ``tail fragmentation'' where only small gaps remain for heavy operators.

    \item \textbf{Greedy Best-Fit Placement:} 
    For each replica $r$, we search for a candidate device $d \in \mathcal{D}$ that minimizes the \textit{residual capacity vector} $\|\mathbf{C}_d - \mathbf{v}_r\|$, where $\mathbf{C}_d$ denotes the current available resources on device $d$. A placement is considered valid only if it satisfies the safety constraint.
    Using the interference model (Eq.~\ref{eq:interference}), we compute the interference-adjusted latency $T'_v = T_v \cdot I_{d,v}(b_v, p_v)$ and verify that the resulting global latency remains within the model’s SLO.
    Only when no existing device can accommodate $r$, a new device is activated, thereby minimizing active GPUs.

    \item \textbf{Locality-Aware Tie-Breaking:}
    {When multiple devices provide comparable fits, \sysname{} prioritizes placements that preserve data locality, minimizing communication overhead.
    It further ranks candidates by communication hierarchy, favoring intra-device, then intra-server (NVLink) or NVL domain, and cross-server (InfiniBand) placements.}
    
\end{enumerate}
The same placement routine is applied incrementally during autoscaling events.
Placement is not accepted independently of provisioning: for every candidate device, \sysname{} feeds the profiled communication cost and interference-adjusted operator latency back into the end-to-end latency model. A candidate that would violate the SLO is rejected; if no existing device remains feasible, \sysname{} activates a new device. Thus, provisioning proposes the logical replica configuration, while placement verifies that the physical realization preserves its SLO estimate.
By avoiding global re-optimization, this greedy strategy maintains low decision latency while providing high-quality placements in dynamic traffic conditions.

\subsection{Execution Plane}
\label{sec:design:runtime}

\sysname{}'s \textit{Execution Plane} actuates the scaling plan generated by the control plane.
To support efficient op-level replicating and execution, we extend the LLM engine runtime with the following novel features and optimizations.

\myparagraph{Dynamic Operator Replicating and Registering}
Existing LLM engines assume a static execution graph with fixed operators, whereas \sysname{} must dynamically adjust operator replicas and execution paths.
To support this, \sysname{} introduces a \textit{Replica Manager} that maintains a dynamic registry of \textit{Operator Replicas}, automatically registering new replicas.
A late-binding mechanism via \texttt{forward\_pre\_hooks} transparently intercepts operator calls, to select the appropriate replica at runtime. This enables in-flight requests to immediately switch to the new execution graph with zero downtime.

Moreover, replicating operators requires allocating GPU memory at runtime while existing LLM engines require all memory to be pre-allocated during engine initialization. To solve this problem, \sysname{} implements an \textit{ElasticBlockManager} by integrating kvcached~\cite{xing2025towards, yu2025prism}. This allows the engine to dynamically allocate memory for newly created replicas and reclaim memory from those removed.

\myparagraph{Multi-Stream Management}
\label{sec:stream_pool}
For a model with $D$ layers, $|\mathcal{V}|$ operators per layer, and up to $R$ replicas per operator, naively creating a unique stream for every instance incurs prohibitive performance overhead due to the frequent switching between streams ($O(D \times |\mathcal{V}| \times R)$). 
We address this bottleneck via a two-tier strategy:
(1) \uline{Stream Reuse} exploits the sequential nature of layers. A \texttt{GlobalStreamPool} recycles streams across pipeline stages. This bounds resource overhead to $O(R)$ regardless of model depth.
(2) \uline{Cross-Device Pipelining} hides the communication latency between consecutive operators. A \texttt{MultiDeviceStreamPool} pipelines remote data transfers (\eg{}, NVLink) with kernel execution, ensuring data arrives exactly when the compute kernel is ready. 
{Each stream is further bound to a green-context resource, with SM allocation derived from the interference model (Eq.~\ref{eq:interference}). This enforces hardware-level partitioning at dispatch time.}

\myparagraph{{Operator Pipelining and Dispatching}}
Monolithic execution runs all operators on a single CUDA stream, creating utilization bubbles as later operators must wait for earlier ones.
To improve throughput, inspired by Nanoflow~\cite{zhu2025nanoflow}, \sysname{} adopts pipelined execution, allowing operators to run concurrently upon available inputs.
{Requests are dispatched to operator replicas via shortest-queue routing, weighted by replica capacity,} and dependent data is synchronized between operators using lightweight CUDA events.

\section{Evaluation}
\label{sec:eval}

We prototyped \sysname{} on top of nano-vLLM~\cite{nanovllm} in~$\sim$17K lines of Python code. nano-vLLM is a minimalist inference framework that inherits key optimizations from vLLM (\eg{}, PagedAttention, continuous batching, and tensor parallelism), ensuring that our results are representative of modern high-performance LLM serving systems.
For a controlled comparison, \emph{all} evaluated approaches use the same nano-vLLM inference engine, including the same model implementation, kernels, batching scheduler, KV-cache manager, and tensor-parallel backend. The experiments reproduce each baseline's autoscaling policy over this common data plane; the intended differences are the scaling policy and granularity, rather than the underlying inference runtime.

\subsection{Experiment Setup}
\label{sec:eval:setup}

\myparagraph{Models and Traces}
We evaluate \sysname{} using two representative LLMs: Qwen2-7B~\cite{qwen2-7b} (a dense LLM) and Qwen2-57B-A14B~\cite{qwenmoe} (an MoE LLM).
These models capture two widely used architectures in generative models.
We use production LLM serving traces~\cite{stojkovic2025dynamollm} which include temporal request arrival patterns and sequence lengths collected from real-world interactions, enabling controlled and reproducible evaluations of autoscaling under dynamic load conditions.

\begin{figure}[t!]
    \centering
    \includegraphics[width=\linewidth]{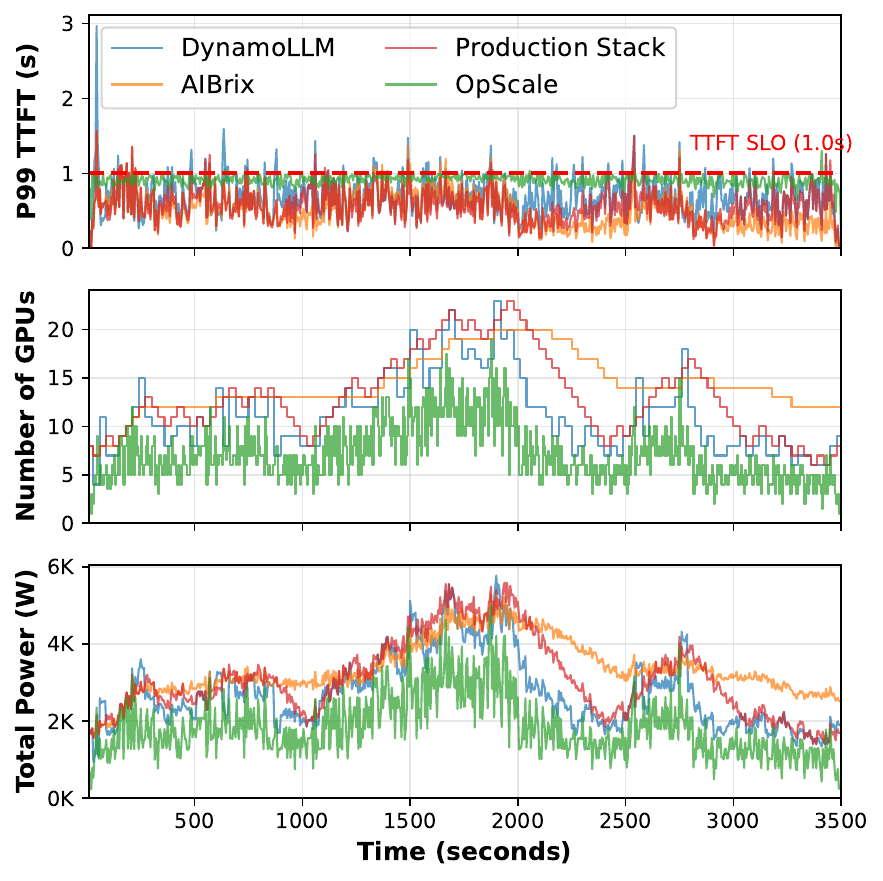}
    \caption{Comparison of request performance, GPU usage, and cluster power usage during autoscaling for Qwen2-7B.
    }
    \label{fig:autoscaling-comparison}
\end{figure}

\myparagraph{Hardware}
Most experiments are conducted on a 40-GPU cluster of five Azure GPU VMs~\cite{a100azure}.
Each VM is equipped with 8 NVIDIA A100-80GB GPUs connected via NVLink and high-speed InfiniBand networking.
For sensitivity study on hardware and granularity, we also evaluate \sysname{} on a cluster of 24 GB200 GPUs within the same NVLink domain.

\myparagraph{Baselines}
We compare \sysname{} against state-of-the-art \textit{model-level} autoscaling systems ported over the common nano-vLLM backend:
(1) DynamoLLM~\cite{stojkovic2025dynamollm} (SLO-driven), which dynamically adjusts model replicas based on max per-instance serving throughput under given TTFT and TBT SLOs.
(2) AIBrix~\cite{team2025aibrix} (utilization mode), which uses GPU utilization metrics to decide scaling actions.
(3) Production Stack~\cite{production-stack} (queue-driven), which triggers scaling based on the number of pending prompt tokens in the queue.
We adopt the recommended configurations in each baseline's documentation. Model scaling frequency is set to every 20 seconds.
Additionally, we add (4) DVFS~\cite{stojkovic2025dynamollm} and $\mu$-Serve~\cite{qiu2024power} for power consumption comparison, and (5) Attn-FFN disaggregation~\cite{wang2025step} for scaling granularity study.

\subsection{Dynamic Autoscaling}
\label{sec:eval:dynamic}
We first evaluate \sysname{} and baselines under dynamic load patterns to assess autoscaling behavior, including GPU usage, request SLO attainment, and total power consumption.
In total, each model serves 929K requests (1.5B prompt tokens) replayed from the production traces~\cite{stojkovic2025dynamollm}.

\Cref{fig:autoscaling-comparison} presents the autoscaling comparison in a one-hour sample of the trace for Qwen2-7B (MoE results are in Appendix~\ref{sec:appendix:more-eval}).
\sysname{} consistently provisions fewer GPUs while delivering the highest SLO attainment across both dense and MoE models: \sysname{} requires on average 7.1 GPUs, a 35--50\% reduction compared to DynamoLLM (11.2), AIBrix (14.3), and Production Stack (13.0).
This is because (1) \sysname{} only scales out critical operators, and (2) it scales down faster to avoid idle provisioning.
Consequently, we observe a reduced power usage (bottom row) from \sysname{}, even though the per-GPU utilization is higher.
In addition, \sysname{}'s faster elasticity (1 second) and fine-grained optimization loop (\S\ref{sec:design:provisioning}) help achieve the best SLO attainment: 98.4\%, compared to 88--95\% for the baselines.

For the MoE model (\Cref{fig:autoscaling:moe}), the trend is similar, but with even higher GPU savings. This aligns with the theoretical analysis (\S\ref{sec:analysis:results}).
\sysname{} allocates only 10.8 GPUs on average, much lower than DynamoLLM (17.3), the GPU-util-based autoscaler (23.1), and the queue-based baseline (18.0).
SLO attainment improves correspondingly: \sysname{} reaches 98.1\%, compared to 84.2--97\% for baselines.
These differences match the behavior visible in the figure --- baselines frequently over-allocate GPUs yet still incur noticeable tail TTFT spikes during fast load surges, whereas \sysname{} maintains stable latency with a substantially smaller footprint.

\myparagraphnodot{Where Do the Savings Come From?}
The savings come from two sources.
First, \sysname{} scales at the operator level rather than at the model level.
The bottleneck operators shift continuously with the workload.
At short sequence lengths, linear project and MLP operators can dominate compute. As sequence length grows, attention becomes increasingly critical because of its steeper sequence-length sensitivity.
\sysname{} tracks this shifting profile and redistributes resources across the full operator set, avoiding the waste of scaling operators that are not the current bottleneck.
Second, scaling itself is fast: sub-second operator actuation keeps the actuation portion of the response to transient spikes small, avoiding the standing overprovisioning that would otherwise be needed to absorb demand while slower scaling catches up.

\begin{figure}[t!]
    \centering
    \begin{minipage}[t]{0.55\linewidth}
        \centering
        \includegraphics[width=\linewidth]{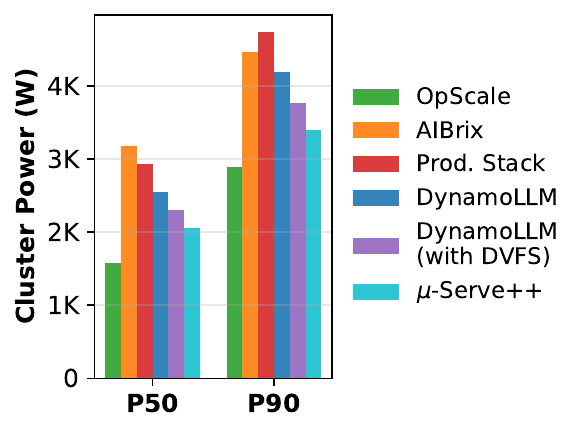}
        \caption{{Cluster-level power consumption comparison.}}
        \label{fig:barplot_power_qwen2}
    \end{minipage}
    \hfill
    \begin{minipage}[t]{0.42\linewidth}
        \centering
        \includegraphics[width=\linewidth]{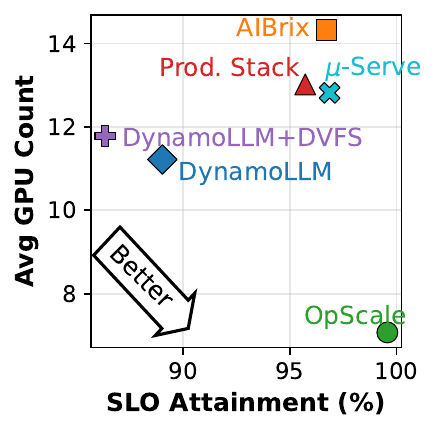}
        \caption{{Cost vs. SLO attainment trade-off.}}
        \label{fig:scatter_cost_vs_slo_qwen2}
    \end{minipage}
\end{figure}

\myparagraphnodot{Impact on Power Consumption.}
{
\Cref{fig:autoscaling-comparison} shows that \sysname{} consistently consumes the lowest cluster-level power (mainly by reducing operator replicas and GPU idle power).

To further evaluate power–performance trade-offs, we compare \sysname{} against two \textit{power-aware} baselines: (1) DynamoLLM+DVFS (enabling GPU frequency scaling) and (2) $\mu$-Serve++ (augmenting $\mu$-Serve’s power-aware model partitioning and placement with DynamoLLM’s autoscaling logic).
\Cref{fig:barplot_power_qwen2} shows that \sysname{} achieves the lowest power consumption at both P50 and P90 (20\% and 16\% less than the best power-aware baseline $\mu$-Serve++).
\Cref{fig:scatter_cost_vs_slo_qwen2} further illustrates the trade-off between performance and cost.
While DVFS-based baselines like DynamoLLM+DVFS and $\mu$-Serve++ improve power consumption, they lag behind \sysname{} in both dimensions of SLO attainment and GPU costs.
These results confirm that structural op-level elasticity is more effective for system-wide power savings than coarse-grained replica scaling or DVFS.
MoE results are in Appendix~\ref{sec:appendix:more-eval}.
}

\begin{table}[!t]
\centering
\caption{Scale-up latency on Qwen2-7B.}
\label{tab:scaling-overhead}
\resizebox{0.97\linewidth}{!}{%
\begin{tabular}{lccc}
\toprule
\textbf{Approach} & \textbf{P99 latency} & \textbf{P90 latency} & \textbf{Avg. latency} \\
\midrule
Model-level        & 11.55 s & 11.04 s & 10.68 s \\
\midrule
OpScale (1 op)     & 0.10 s & 0.05 s & 0.03 s \\
OpScale (50\% ops) & 0.42 s & 0.26 s & 0.18 s \\
OpScale (all ops)  & 0.45 s & 0.38 s & 0.33 s \\
\bottomrule
\end{tabular}}
\end{table}

\myparagraphnodot{Scale-up Latency.}
{A major contribution to better SLO attainment under dynamic traffic is the rapid response to the demand changes.}
To understand elasticity limits, we compare the scaling overhead of op-level provisioning against traditional model-level scaling across a range of scaling intensities.
\Cref{tab:scaling-overhead} reports results on Qwen2-7B.
Model-level scaling incurs non-trivial per-model-instance latency of 10.7 seconds on average, as each scale-out event requires (1) loading the full set of model weights and (2) provisioning an additional LLM engine control plane.
On the other hand, \sysname{} needs to load and initialize replicas only for the provisioned operators.
In addition, it avoids replicating the LLM engine control plane, which together reduces scaling overhead by up to two orders of magnitude (P99 <0.45\,s), even when provisioning many operators simultaneously.
A fully warm standby instance can reduce this startup latency, but it must reserve enough GPU memory and compute capacity for an entire model replica even while idle.
Under our cost objective, such reserved GPUs are counted as provisioned capacity; a warm standby therefore exchanges persistent over-provisioning for faster response rather than eliminating the scaling cost.
\Cref{tab:scaling-overhead} instead compares on-demand scale-out without pre-reserving an additional full-model replica.

\subsection{Cost Savings when Meeting SLOs}
\label{sec:eval:cost-savings}

\begin{figure}[!t]
  \centering
  \begin{subfigure}[b]{0.47\textwidth}
    \centering
    \includegraphics[width=1\textwidth]{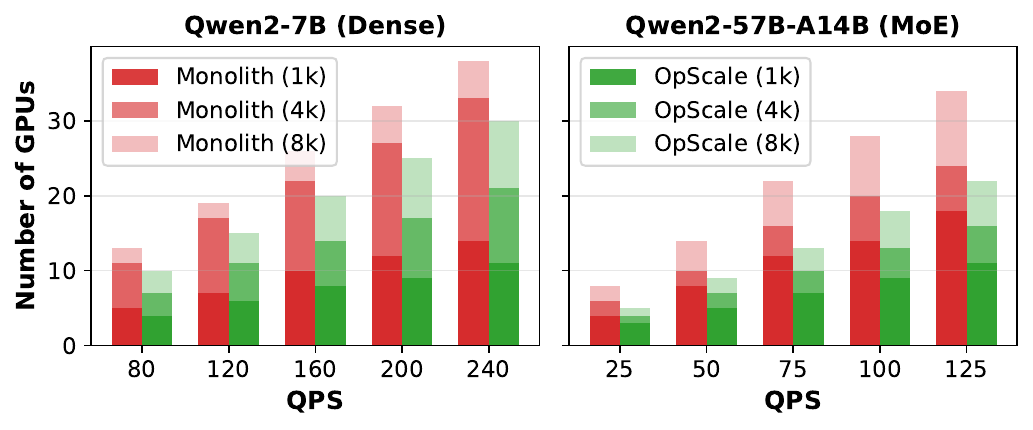}
    \caption{Number of GPUs required.}
    \label{fig:cost-gpus}
  \end{subfigure}
  \begin{subfigure}[b]{0.47\textwidth}
    \centering
    \includegraphics[width=\textwidth]{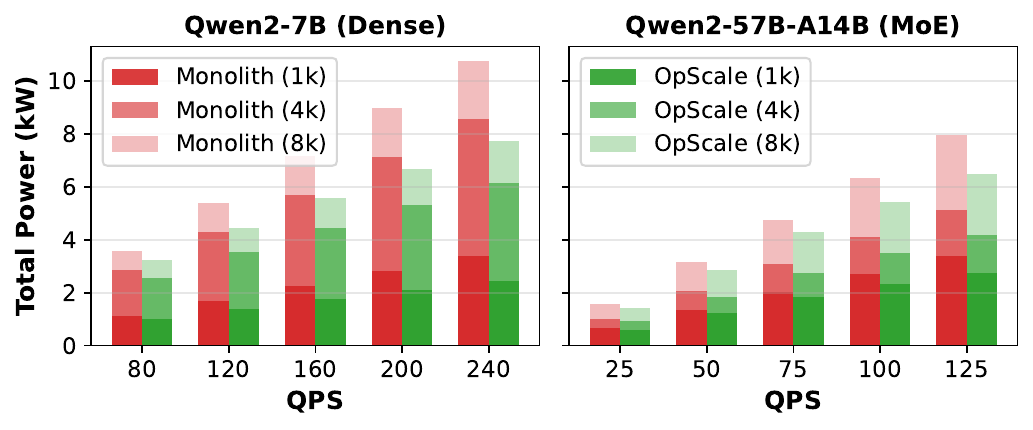}
    \caption{Average total power consumption.}
    \label{fig:cost-power}
  \end{subfigure}%
  \caption{Comparison of the serving cost (in total GPUs and total power consumption) for varying QPS.}
  \label{fig:cost}
\end{figure}

To understand the limits of cost savings, we evaluate the minimum capacity required to sustain varying workloads, while meeting the same latency SLOs.
For each QPS level, we configure and scale both systems (model-level vs. \sysname{}) until the SLO violation rate falls below the threshold (\eg{}, when P99 TTFT of Qwen2-7B is smaller than one second).

\Cref{fig:cost} presents the provisioned capacity, measured in total GPUs and total power draw, required to sustain increasing traffic in QPS. Across both dense and MoE models, \sysname{} consistently achieves the target SLO with 20.1\% and 35.7\% fewer GPUs on average, respectively, for requests with 1K sequence lengths.
This reduction stems from \sysname{}'s fine-grained op-level provisioning, which avoids the over-provisioning required by monolithic model-level provisioning.
\sysname{}'s savings for longer requests vary from 36.3\% (for 4K) to 22.1\% (for 8K) under the same SLO target, aligning with our observation from the theoretical analysis (\S\ref{sec:analysis:results}).

The reduction in GPU usage directly translates into lower power consumption (\Cref{fig:cost-power}).
At high load, \sysname{} reduces cluster-wide power draw by 14--28\% (depending on the model and sequence length), even though the per-GPU power can be higher in \sysname{} (\eg{}, 282 W vs. 245 W on average).
Notably, unlike baselines whose power usage grows sharply with QPS due to coarse scaling actions in rigid GPU increments, \sysname{} maintains a smoother growth curve, reflecting its ability to exploit latent compute slack at the operator level before allocating additional hardware.

\subsection{Throughput Improvements}
\label{sec:eval:static-throughput}

\begin{figure}[!t]
    \centering
    \includegraphics[width=\linewidth]{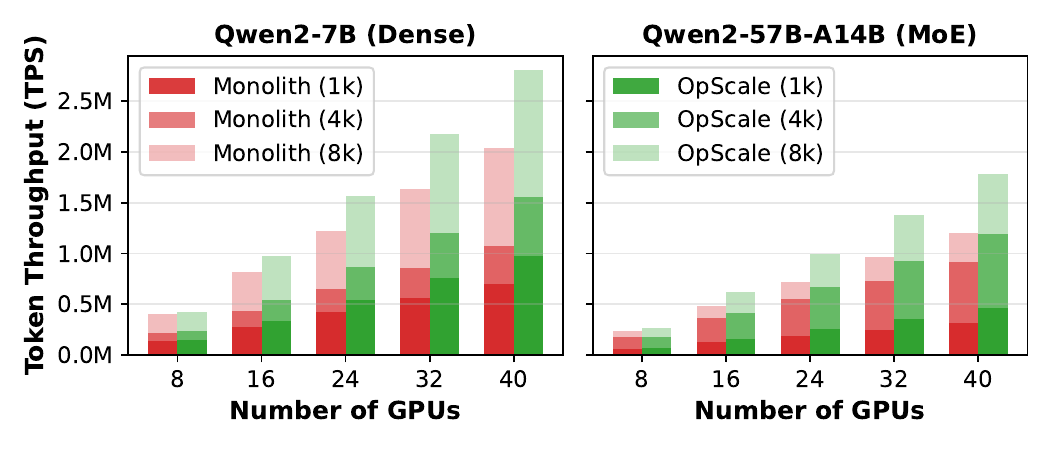}
    \caption{Max throughput achieved at fixed provisioning.}
    \label{fig:throughput}
\end{figure}

Op-level provisioning also benefits \textit{static} deployments, with a fixed number of GPUs.
In static clusters up to 40 GPUs, we increase the request arrival load until each system reaches its SLO limit and report the maximum sustainable input token throughput (TPS).
Shown in \Cref{fig:throughput}, \sysname{} achieves higher throughput from the same hardware allocation.
For the dense Qwen2-7B model, \sysname{} improves peak TPS by 3--38\% across cluster sizes and request sequence lengths.
The improvement is even larger for the MoE model, where \sysname{} achieves up to 44\% higher TPS at 40 GPUs.
These improvements follow the same mechanism as the dynamic scaling results: the fixed budget is directed toward the operators that limit throughput.
At shorter contexts this favors the heavy linear or fused expert operators, while longer contexts shift capacity toward attention; lightweight operators retain fewer replicas and are packed into the residual SM/memory capacity.
Model-level provisioning cannot make this exchange because each added replica duplicates every operator.

\begin{figure}[!t]
    \centering
    \includegraphics[width=0.98\linewidth]{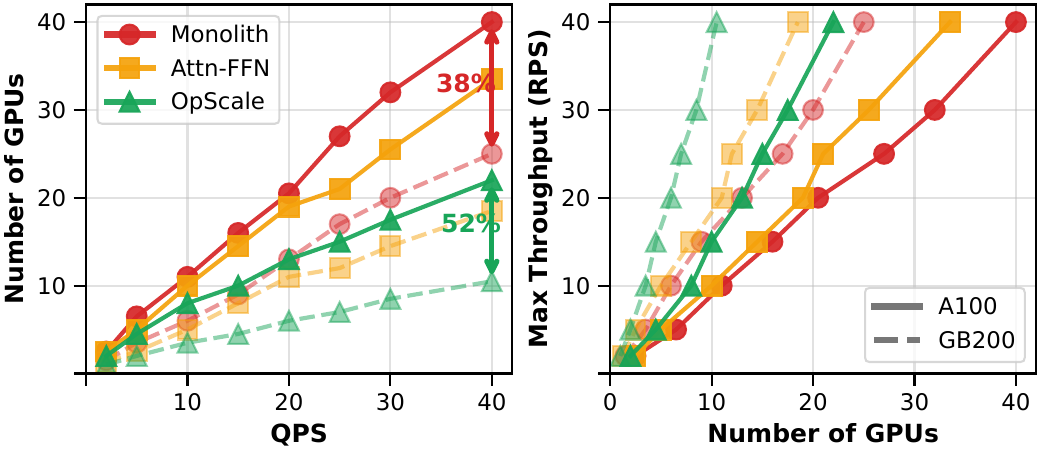}
    \caption{Static provisioning comparison across serving granularities and hardware configurations.}
    \label{fig:attn-ffn}
\end{figure}

\myparagraphnodot{Beyond Attn-FFN.}
{
We further compare \sysname{} against Attn-FFN partitioning~\cite{wang2025step,zuo2025serving,zhu2025megascale}, which only decouples attention and FFN blocks (\Cref{fig:attn-ffn}).
While Attn-FFN reduces GPU demand compared to model-level baseline, it is consistently outperformed by \sysname{}'s full operator-level elasticity.
On A100 clusters, \sysname{} reduces GPU demand by up to 33\%, translating to a 1.7$\times$ throughput gain.
The advantage of \sysname{} scales with hardware interconnect.
When switching from A100 cluster to GB200~\cite{gb200azure} (with NVL domain), \sysname{} reduces the required GPU count by 52\% at peak load while the benefits drop to 38\% for the baseline.
With op-level decoupled provisioning, \sysname{} better exploits high-speed interconnects to avoid the resource fragmentation inherent in coarser block-level or replica-level approaches.}

Overall, even without elastic scaling, \sysname{} can drive significantly higher throughput from the same provisioned devices, demonstrating improved efficiency under \textit{both} dynamic (\S\ref{sec:eval:dynamic}) and static (\S\ref{sec:eval:static-throughput}) deployments.

\subsection{Robustness and Sensitivity Analysis}
\label{sec:eval:sensitivity}

\begin{figure}[!t]
    \centering
    \includegraphics[width=\linewidth]{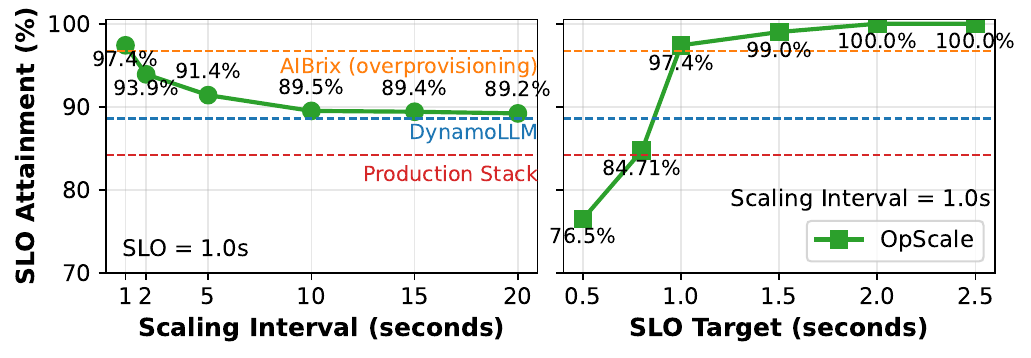}
    \caption{(Left) Impact of the scaling interval on SLO attainment. (Right) Impact of SLO target. Both with Qwen2-7B.}
    \label{fig:sensitivity}
\end{figure}

We analyze how \sysname{} behaves under different autoscaling intervals and SLO targets, to characterize its robustness.
Shown in \Cref{fig:sensitivity} (left), the SLO attainment rate increases as the autoscaling interval decreases within the 1--20 seconds range, reaching 97.4\% when the interval is set to one second.
This aligns with our observation from serving production LLM workloads (\Cref{fig:scaling-overhead-trend}), where a shorter scaling interval allows the system to react before the traffic deviates significantly, reducing queue buildup and improving SLO attainment.
Horizontal lines indicate model-level approaches scaling every 20 seconds.
At a 20-second interval, \sysname{}'s SLO attainment drops to 89\%, still surpassing the model-level baselines. The only exception is AIBrix, which unfortunately overprovisions with the utilization-based scaling mode.

In \Cref{fig:sensitivity} (right), we fix the autoscaling interval to one second and look at the SLO attainment rates under varying SLO targets.
SLO target for the baselines (dashed lines) is one second.
\sysname{} achieves 100\% SLO attainment when the SLO target is relaxed, but the benefits of op-level autoscaling would diminish.
On the other hand, even with a stricter SLO (0.8 seconds), \sysname{} achieves the same SLO attainment, compared to Production Stack with an SLO of one second.

Additional sensitivity analysis on model sizes, prefill/decode stages, sequence lengths are in \S\ref{sec:analysis:results} and Appendix~\ref{sec:appendix:prefill_vs_decode}.

\subsection{\sysname{} Overhead}
\label{sec:eval:overhead}

\myparagraphnodot{Online Overhead and Scalability.}
{
\sysname{} imposes negligible control-plane overhead.
For Qwen2-7B, the optimizer computes a full scaling plan in 2.6\,ms median (P99 3.2\,ms), including 2.5\,ms for provisioning and 0.1\,ms for placement, all on CPU.
This provides $\sim$370$\times$ headroom in a 1-sec scaling interval. For the larger Qwen2-57B-A14B model, total time remains low at 4.4\,ms. 
At the execution plane, overhead consists of (1) per-operator load-balancing lookups and (2) per-replica dispatch costs. This contributes a 0.3\% total overhead (\ie{}, $\sim$1.5\,ms out of a $\sim$500-ms prefill iteration).
We mitigate dispatch overhead by replacing blocking synchronizations with event-based ordering.
With lazy-sync optimization and pipelining the remaining 90\,$\mu$s of CPU work behind GPU kernel execution, \sysname{} successfully hides architectural overhead, ensuring total overhead in the execution plane remains negligible relative to end-to-end performance.
}

\myparagraphnodot{Offline Profiling Overhead.}
{
Operator profiling is a one-time, offline effort.
With sparse sampling and structural equivalence across layers (\S\ref{sec:design:profiling}), profiling a 57B model can complete in under an hour, on a single GB200 node.
These costs are amortized across model generations; because successive releases within a model family typically retain core operator types (\eg{}, Attention and MLP with identical kernel implementations), profiles collected for one generation largely transfer to the next---only operators whose kernel structure has changed (\eg{}, a switch from multi-head to grouped-query attention) require re-profiling.
For example, the Llama~2$\to$3 and Qwen~1.5$\to$2 transitions preserve the same SwiGLU MLP and RoPE attention operators; only the attention head grouping and vocabulary-embedding dimensions differ, limiting re-profiling to a small subset of operators.
Cross-family reuse is also feasible when models share common operator implementations (\eg{}, FlashAttention), further reducing the marginal profiling cost for new models.
}

\begin{table}[t]
\centering
\caption{Accuracy of profiling-driven models (relative error).
}
\label{tab:model-accuracy}
\resizebox{0.95\linewidth}{!}{%
\begin{tabular}{lccc}
\toprule
\textbf{Component} & \textbf{Error Rate Metric} & \textbf{Average} & \textbf{P90} \\
\midrule
Operator profile & Sensitivity prediction & 7\% & 15\% \\
SM contention model  & Slowdown prediction & 5\% & 9.4\% \\
Queueing model       & Latency prediction & 0.8\% & 1.9\% \\
\bottomrule
\end{tabular}}
\end{table}

\subsection{Profiling-driven Model Accuracy}
\label{sec:eval:prediction-accuracy}
{
We evaluate the accuracy of the underlying models that drive \sysname{}'s provisioning and placement decisions. Summary results are reported in \Cref{tab:model-accuracy}.

\myparagraph{Operator Sensitivity Modeling}
We compare predicted performance from our piecewise sensitivity model (\S\ref{sec:design:profiling}) against measured runtime metrics. The average relative error is 7\% (P90 15\%). Since operator provisioning depends primarily on identifying \emph{relative bottlenecks} rather than absolute latency values, this fidelity is sufficient to guide optimal scaling while significantly reducing profiling effort.

\myparagraph{SM Contention Modeling}
\sysname{} models SM contention when operators are colocated on the same GPU.
Unlike prior interference models~\cite{dhakal2020gslice,gpulet,shubha2024usher} that struggle to generalize~\cite{huang2026mushare}, \sysname{}'s empirical degradation model (\S\ref{sec:design:placement}) estimates latency inflation as a function of allocated SM fractions.
During runtime validation, the average slowdown prediction error is 5\% (P90 9.4\%). This accuracy enables effective colocation and achieves substantial throughput gains (\S\ref{sec:eval:static-throughput}) by balancing spatial and temporal GPU resources.
While our approach operates at the SM allocation granularity, incorporating finer-grained kernel-level spatial and temporal multiplexing (\eg{}, LithOS~\cite{coppock2025lithos}) could further reduce interference and improve accuracy, which we leave to future work.

\myparagraph{Queue Modeling}
The analytical networked queueing model underlying the control plane (\S\ref{sec:analysis}) estimates the end-to-end iteration latency by composing per-operator sojourn times.
Across all evaluated scenarios and bursty trace replays, the average relative error is 0.8\% (P90 1.9\%).
This high fidelity demonstrates that the queueing network abstraction of operators is robust to realistic workload variability.
}
\section{Discussion}
\label{sec:eval:discussion}

\myparagraphnodot{When to Adopt Op-Level Autoscaling?}
Op-level autoscaling is most beneficial when workloads do not require \textit{ultra-low} latency (which is common, and growing agentic applications have even more relaxed SLOs~\cite{pan2025measuring}) and models continue to exhibit heterogeneous operator behaviors (as shown in \S\ref{sec:bg:characterization}).
These two factors present opportunities of model deployment flexibility, in order to achieve resource efficiency.
In contrast, inference systems pursuing ultra-low latency through megakernel~\cite{megakernel,cheng2025mirage} (fusing all operators into a single large kernel) leave limited room for op-level provisioning and scaling.

\myparagraphnodot{Prefill-Decode Disaggregation or Co-location?}
Our analysis (\S\ref{sec:appendix:prefill_vs_decode}) shows that prefill stages, being compute-intensive and variable in demand, gain the most from fine-grained autoscaling.
Decode, while often constrained by tighter latency budgets, benefits less from op-level provisioning.
This suggests that enabling prefill-decode disaggregation complements op-level scaling by exposing varying extents of optimization opportunities across stages.
In prefill-decode co-location cases with chunked prefill~\cite{agrawal2024medha,sarathi-serve}, the maximum sequence length is bounded by the chunk size, altering operator load distribution.
Op-level autoscaling remains valuable in such cases, as it can adapt provisioning to dynamic chunking~\cite{agrawal2024medha}.

\myparagraphnodot{Op-level Resharding.}
{
Although \sysname{} supports dynamic op-level resharding (changing tensor parallelism) online (\S\ref{sec:analysis:formulation}), we observe that horizontal scaling (\eg{}, scaling replicas) consistently delivers higher throughput-per-GPU gains across traces.
In addition, due to weight redistribution and synchronization costs across GPUs, op-level resharding incurs 11$\times$ higher overhead (P99 1.15s) than replica scaling.}

\myparagraphnodot{Multi-Tenant, Multi-Model Serving.}
\sysname{} focuses on single-model serving, where a dedicated set of GPUs is provisioned for one model---the predominant deployment pattern in production LLM clusters~\cite{stojkovic2025dynamollm,qin2024mooncake,qiu2025modserve}. However, it is complementary to multi-model and multi-tenant GPU sharing~\cite{duan2024muxserve,li2023alpaserve,xiang2025aegaeon,xing2025towards,yu2025prism,qiao2025conservefinegrainedgpuharvesting}.
First, as \sysname{} manages individual operators rather than monolithic model replicas, it can pack operators from \textit{different} models onto shared GPUs with its contention-aware placement algorithm (\S\ref{sec:design:placement}).
Second, scaling down non-critical operators of one model frees SM/memory that can be immediately reused by operators of another model. This degree of reuse is beyond model-level scaling.
Extending to multi-tenant settings with request-level multiplexing across models requires tenant fairness policies and cross-model interference modeling, which we leave to future work.

\myparagraphnodot{Autoscaling.}
Autoscaling has been a crucial technique for system operations, even before the emergence of LLM serving. One system with such extensive efforts is microservices~\cite{qiu2020firm, gias2019atom, wang2022deepscaling, wang2024autothrottle}.
While the overall concept of autoscaling remains, LLMs require us to revisit prior assumptions.
For example, while CPU usage alone is a strong signal for autoscaling microservices in production~\cite{k8s_hpa,wang2022deepscaling}, LLM inference is strongly sensitive to workload characteristics and GPU resources.
To this end, {\sysname} takes the first step to formal characterization of LLM inference at op-level granularity, and addresses provisioning and scaling challenges in practice.

\section{Related Work}

\myparagraph{Disaggregation in LLM Serving}
There has been a trend of disaggregation on generative model serving:
(1) Prefill-Decode disaggregation~\cite{patel2024splitwise,zhong2024distserve}, enabling independent autoscaling of each stage.
(2) Encoder–LLM disaggregation~\cite{qiu2025modserve,epd,dong2025hydrainfer} and VAE-DiT disaggregation~\cite{huang2025ddit} focus on modality-specific scaling.
(3) MA parallelism~\cite{wang2025step,zuo2025serving,zhu2025megascale,chowdhery2023palm,liu2025expert} disaggregates attention and expert/FFN layers to leverage pipelining and improve throughput.
\sysname{} differs from these fixed decompositions in both objective and mechanism: it identifies workload-dependent bottleneck operators, independently adjusts their capacity, and jointly places the resulting replicas under communication and interference constraints.
Thus, the contribution is not decomposition alone, but using operator-level elasticity as a cluster-wide autoscaling and placement mechanism.

\myparagraph{LLM Autoscaling Policy}
As mentioned in \S\ref{sec:bg}, AIBrix~\cite{team2025aibrix}, DynamoLLM~\cite{stojkovic2025dynamollm}, Chiron~\cite{patke2025hierarchical}, DeepServe~\cite{deepserve}, and SageServe~\cite{sageserve} propose model-level autoscaling \textit{policies} that combine demand prediction with replica management.
Our work is orthogonal: rather than policy design, we contribute a new autoscaling \textit{mechanism} at the operator level, which benefits from advanced autoscaling policies in this field.

\myparagraph{Multi-Stream Scheduling}
Recent work~\cite{lin2025bullet,zhu2025nanoflow,kamath2025pod,chen2026towards,hong2025semi} has leveraged multi-stream processing to improve utilization and throughput, distinct from our focus on autoscaling and fine-grained resource management.
Nanoflow~\cite{zhu2025nanoflow} exploits intra-device parallelism by overlapping computation with I/O, while Pod-Attention~\cite{kamath2025pod} collocates prefill and decode stages to saturate compute and memory bandwidth.
Bullet~\cite{lin2025bullet} further utilizes multi-stream processing for higher utilization.
These approaches focus on optimizing the \emph{execution pipeline} of a single model instance. 
\sysname{} is orthogonal to these techniques by exploring the optimal serving granularity and how each operator should be provisioned and dynamically scaled across the GPU cluster to minimize serving costs.

\myparagraph{Resource-Level Multiplexing}
{
Several systems optimize model execution by fine-grained decomposition for better resource utilization.
Orion~\cite{strati2024orion} and LithOS~\cite{coppock2025lithos} focus on kernel-level spatial and temporal multiplexing to improve intra-GPU utilization in \textit{multi-tenant} environments.
Similarly, $\mu$-Serve~\cite{qiu2024power} exploits operator sensitivity to GPU frequency to optimize power via DVFS. 
While these efforts operate at a fine granularity, they focus on either SM allocation or frequency scaling on each GPU device.
In contrast, \sysname{} is the first op-level LLM inference framework that addresses \emph{system-level elasticity}, dynamically determining the optimal operator provisioning and scaling plan at a cluster scale.
}
\section{Conclusion}

\sysname{} rethinks serving granularity by shifting from monolithic replicas to operator-level autoscaling. Exploiting operator heterogeneity enables fine-grained resource provisioning, reducing scaling latency to sub-second scales. On production traces, \sysname{} maintains strict SLOs, while reducing GPU allocation by up to 36.3\% and power usage by up to 28\%.

\begingroup
\raggedright
\bibliographystyle{plain}
\bibliography{references}
\endgroup

\appendix

\section{Complementary Analysis Figures}

\begin{figure*}[!t]
    \centering
    \includegraphics[width=\textwidth]{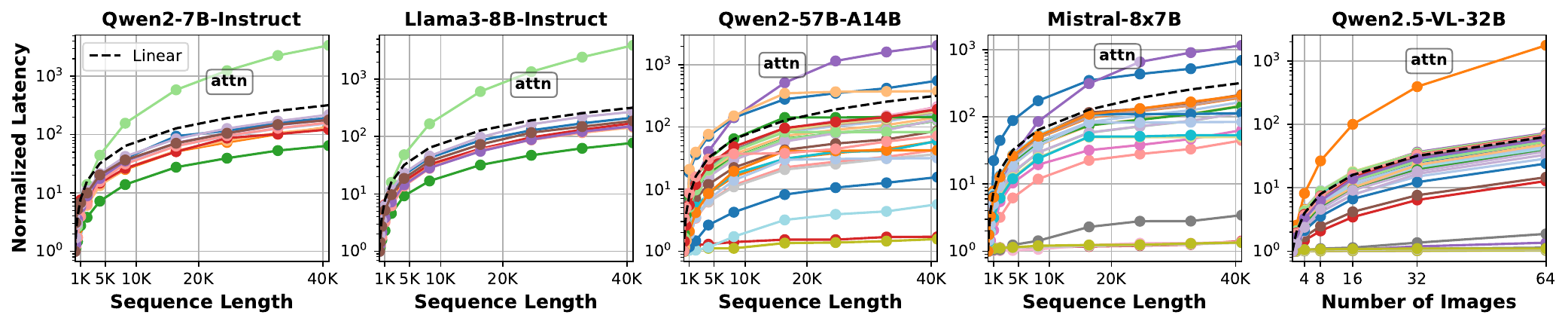}
    \caption{Compute sensitivities to input data size, for various operators in different model architectures.}
  \label{fig:comp-sensitivity}
\end{figure*}

\begin{figure*}[!t]
    \centering
    \includegraphics[width=\textwidth]{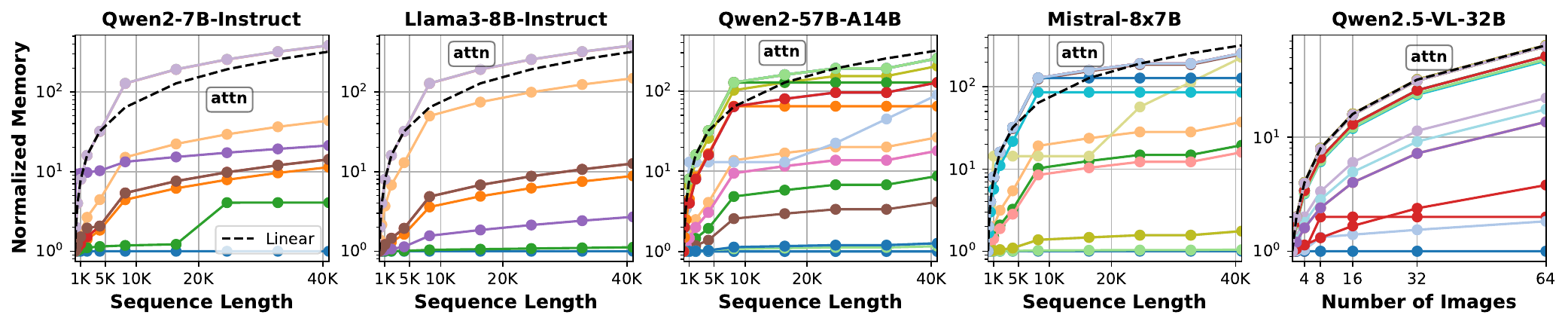}
    \caption{Memory sensitivities to input data size, for various operators in different model architectures.}
  \label{fig:mem-sensitivity}
\end{figure*}

\begin{figure}[!t]
  \centering
  \begin{subfigure}[b]{0.5\textwidth}
    \centering
    \includegraphics[width=1\textwidth]{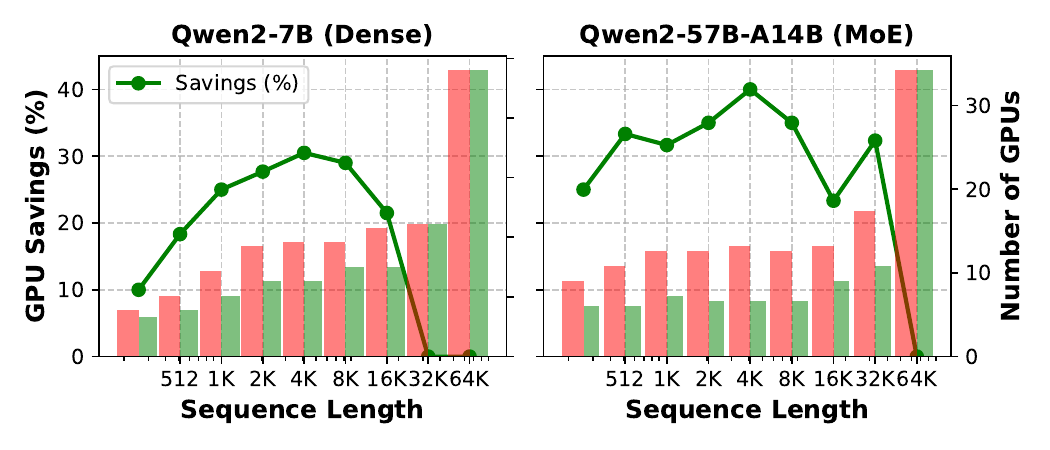}
    \caption{GPU savings.}
    \label{fig:analysis-seqlen-gpus}
  \end{subfigure}
  \begin{subfigure}[b]{0.5\textwidth}
    \centering
    \includegraphics[width=\textwidth]{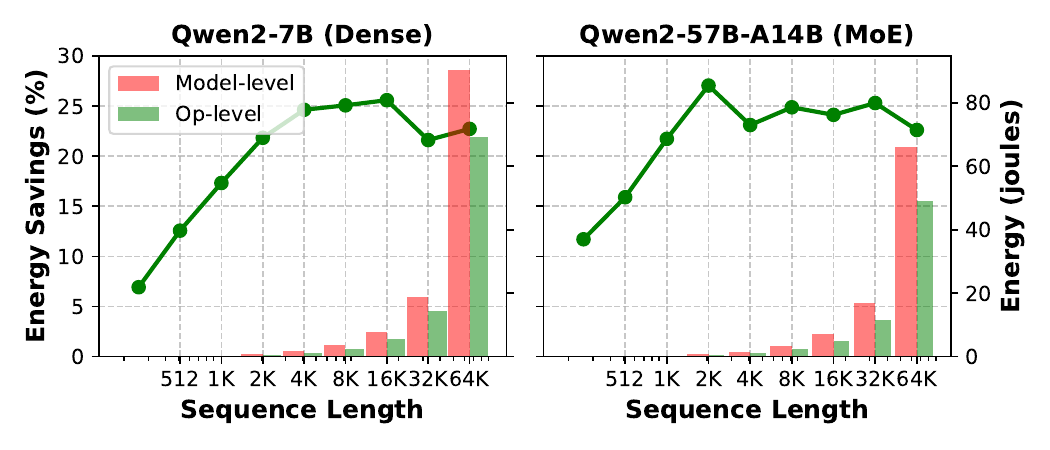}
    \caption{Energy savings.}
    \label{fig:analysis-seqlen-energy}
  \end{subfigure}
  \begin{subfigure}[b]{0.5\textwidth}
    \centering
    \includegraphics[width=\textwidth]{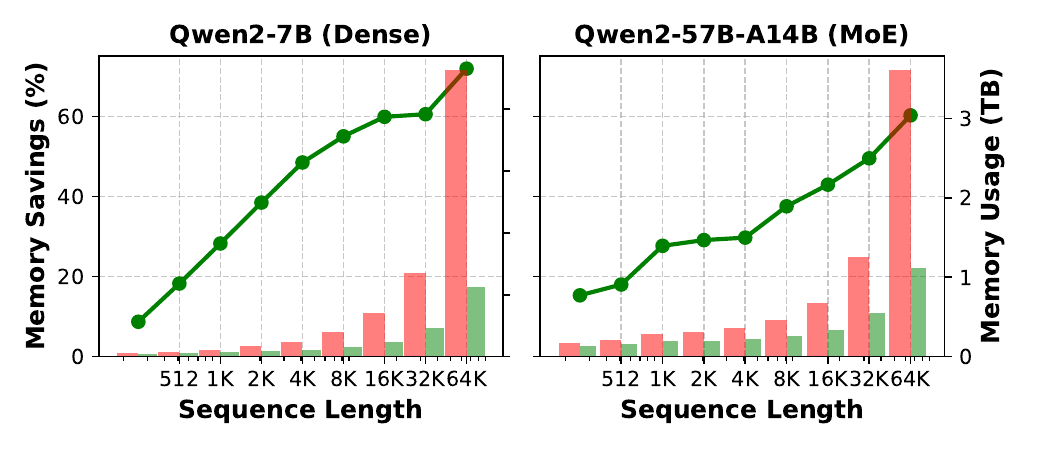}
    \caption{Memory savings.}
    \label{fig:analysis-seqlen-memory}
  \end{subfigure}%
  \caption{Benefits of operator-level resource management under varying sequence lengths. Left Y-axis denotes saving percentages. Right Y-axis denotes the absolute GPU usage, energy consumption, and memory footprint.}
  \label{fig:analysis-seqlen}
\end{figure}
\begin{figure}[!t]
  \centering
  \begin{subfigure}[b]{0.5\textwidth}
    \centering
    \includegraphics[width=1\textwidth]{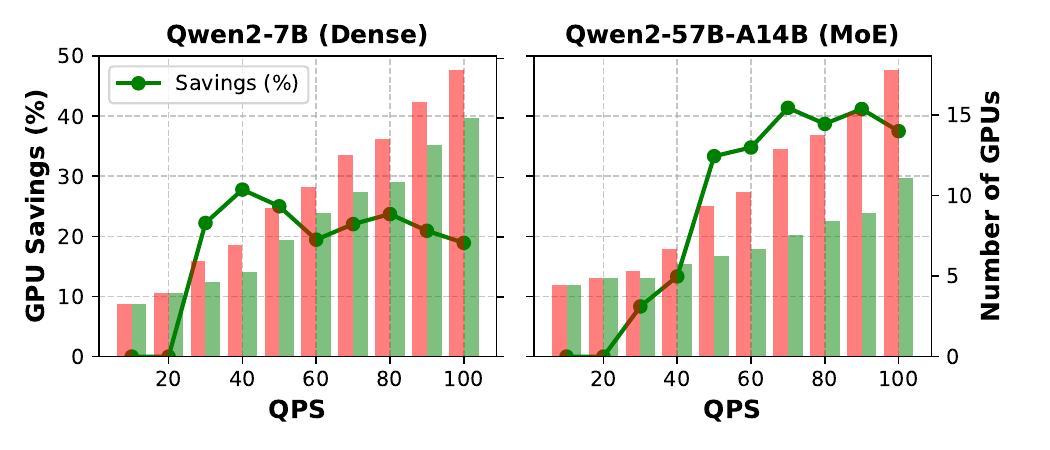}
    \caption{GPU savings.}
    \label{fig:analysis-qps-gpus}
  \end{subfigure}
  \begin{subfigure}[b]{0.5\textwidth}
    \centering
    \includegraphics[width=\textwidth]{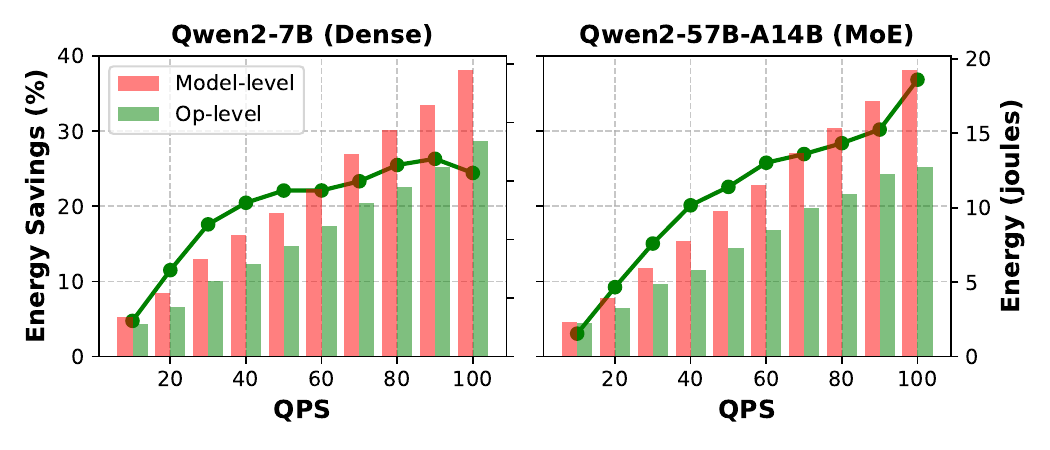}
    \caption{Energy savings.}
    \label{fig:analysis-qps-energy}
  \end{subfigure}
  \begin{subfigure}[b]{0.5\textwidth}
    \centering
    \includegraphics[width=\textwidth]{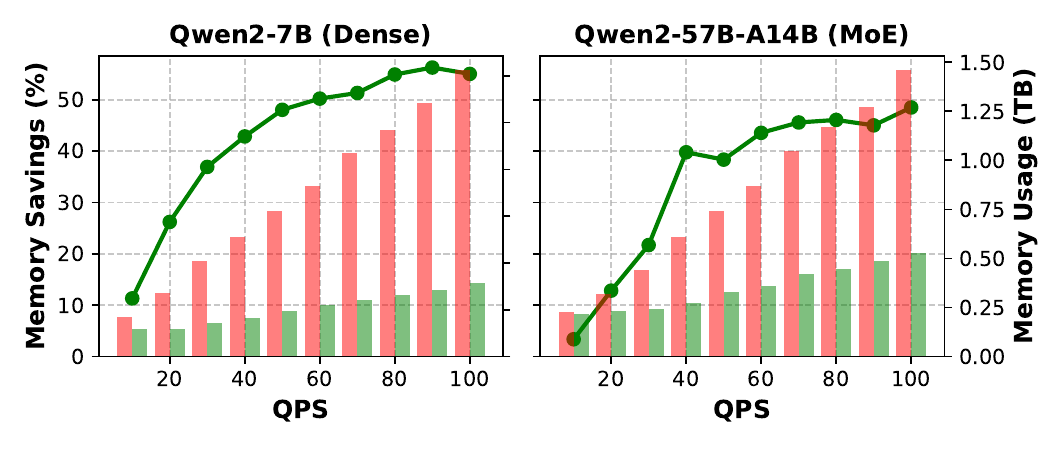}
    \caption{Memory savings.}
    \label{fig:analysis-qps-memory}
  \end{subfigure}%
  \caption{Benefits of operator-level resource management under varying QPS. Left Y-axis denotes saving percentages. Right Y-axis denotes the absolute GPU usage, energy consumption, and memory footprint.}
  \label{fig:analysis-qps}
\end{figure}

\Cref{fig:comp-sensitivity} and \Cref{fig:mem-sensitivity} show details of operator compute and memory sensitivities to sequence length across different model architectures (complementary to \Cref{fig:op-comp-mem-sensitivity}).
\Cref{fig:analysis-seqlen} separately plots GPU savings, energy savings, and memory savings for op-level autoscaling under varying sequence lengths. \Cref{fig:analysis-qps} separately plots GPU savings, energy savings, and memory savings for op-level autoscaling under varying QPS (complementary to \Cref{fig:analysis-seqlen-combined} and \ref{fig:analysis-qps-combined}).

\section{Algorithm Pseudocode}
\label{sec:appendix:analysis}

\begin{algorithm}[t!]
\caption{Greedy Operator Provisioning Algorithm}
\label{alg:greedy-autoscale}
\begin{algorithmic}[1]
\Require DAG $\mathcal{G}=(\mathcal{V},\mathcal{E})$, QPS $q$, batch limits $B_v^{\max}$, parallelism sets $\mathcal{P}_v$, latency functions $T_v(b,p)$ from profiling, SLO $T_{slo}$ with a buffer $\varepsilon$
\State Initialize $p_v \gets \min \mathcal{P}_v$ and $b_v \gets 1$ for all $v\in \mathcal{V}$
\ForAll{$v\in \mathcal{V}$} \Comment{Per-operator initialization}
    \State $\lambda \gets \textsc{ArrivalRates}(\mathcal{G}, q, b_v)$
    \State $r_v \gets \left\lceil \lambda / \mu_v(1,p_v) \right\rceil$ \quad where $\mu_v(b,p)=b/T_v(b,p)$
    \State $(b_v,r_v) \gets \arg\min\limits_{b \in \{1,\ldots,B_v^{\max}\}} \; s_v\!\left(\lambda,\left\lceil \lambda/\mu_v(b,p_v)\right\rceil,b,p_v\right)$
\EndFor
\State Recompute $\lambda \gets \textsc{ArrivalRates}(\mathcal{G}, q, \{b_v\})$
\State $s_v \gets W_v(\lambda_v,r_v,\mu_v(b_v,p_v)) + T_v(b_v,p_v)/b_v$ for all $v$
\State $T \gets \textsc{CriticalPathLatency}(\mathcal{G},\{s_v\})$
\While{true}
    \If{$T \le T_{slo}-\varepsilon$} \Comment{Scaling down}
        \State $j \gets \textsc{BottleneckOnCriticalPath}(\mathcal{G},\{s_v\})$
        \State $\mathcal{M} \gets \{(r_j-1,b_j,p_j)\} \cup \{(r_j-1,b,p_j)\mid b\in[b_j,B_j^{\max}]\} \cup \{(r_j-1,b,p)\mid b\in[b_j,B_j^{\max}],\, p\in\mathcal{P}_j\}$
        \State Filter $\mathcal{M}$ with stability check $\lambda_j < (r_j')\,\mu_j(b',p')$
        \State For each $m\in\mathcal{M}$: tentatively recompute $T'$
        \State Choose $m^\star \in \arg\min\{ \textsc{Cost}(\mathbf{r}',\mathbf{p}') \mid T' \le T_{slo} \}$
        \If{$m^\star$ exists} apply $m^\star$, set $T\gets T'$, continue
        \Else \, \textbf{break} \Comment{No further safe downscale}
    \EndIf
    \ElsIf{$T > T_{slo}$} \Comment{Scaling up}
        \State $j \gets \textsc{BottleneckOnCriticalPath}(\mathcal{G},\{s_v\})$
        \State $\mathcal{M} \gets \{(r_j+1,b_j,p_j)\} \cup \{(r_j+1,b,p_j)\mid b\in [1,B_j^{\max}]\} \cup \{(r_j+1,b,p)\mid b\in[1,B_j^{\max}],\, p\in\mathcal{P}_j\}$
        \State For each $m\in\mathcal{M}$: tentatively re-evaluate $T'$
        \State Choose $m^\star \in \arg\max\{T-T'\}$; Prefer the smallest $\Delta r_j$ that achieves $T' \le T_{slo}$
        \If{$m^\star$ exists} apply $m^\star$, set $T\gets T'$, continue
        \Else \, \textbf{break} \Comment{Cannot improve further}
        \EndIf
    \Else
        \, \textbf{break} \Comment{Within tolerance of SLO}
    \EndIf
\EndWhile
\State \textbf{return} $\{(r_v,b_v,p_v)\}_{v\in \mathcal{V}}$
\end{algorithmic}
\end{algorithm}

\begin{algorithm}[t!]
\caption{Greedy Op-Level Placement}
\label{alg:greedy-placement}
\begin{algorithmic}[1]
\Require DAG $\mathcal{G}=(\mathcal{V},\mathcal{E})$, config $\{(r_v,b_v,p_v)\}_{v\in\mathcal{V}}$ from Alg.~\ref{alg:greedy-autoscale}, device set $\mathcal{D}$, device capacities $\{M_d,U_d\}_{d\in\mathcal{D}}$, interference model $I_{d,v}(b,p)\ge 1$ (from profiling).
\State $k_{\text{base}} \leftarrow \min_{v\in\mathcal{V}} r_v$ \Comment{Number of full model instances}
\State Construct replica sets:
\State \quad $\mathcal{R}_{\text{base}} \leftarrow \{(v,i)\mid v\in\mathcal{V},\; i\in[1,k_{\text{base}}]\}$
\State \quad $\mathcal{R}_{\text{extra}} \leftarrow \{(v,i)\mid v\in\mathcal{V},\; i\in[k_{\text{base}}+1,\,r_v]\}$
\State Sort $(v,k) \in \mathcal{R}_{\text{extra}}$ in descending order of $T_v$
\State $\mathcal{D}_{\text{base}} \leftarrow \textsc{DeployModelInstance}(\mathcal{R}_{\text{base}})$
\State $\mathcal{D}_{\text{empty}} \gets \mathcal{D} \setminus \mathcal{D}_{\text{base}}$
\ForAll{$(v,k)\in\mathcal{R}_{\text{extra}}$}
    \State $Candidates \gets \emptyset$
    \ForAll{$d\in\mathcal{D}_{\text{base}}$} \Comment{Try existing devices first}
        \If{$MemLoad_d + M_v > M_d$} \State \textbf{continue} \EndIf
        \State $T_v' \gets T_v\cdot I_{d,v}(b_v,p_v)$
        \If{$\textsc{ReComputeLatency}(\mathcal{G}) > T_{slo}$} \State \textbf{continue} \EndIf
        \State $slack\_mem \gets M_d - (MemLoad_d + M_v)$
        \State $slack\_comp \gets U_d - (CompLoad_d + T_v')$
        \State $Candidates.\textsc{Append}(d)$
    \EndFor
    \If{$Candidates == \emptyset$} \Comment{No existing device fits}
        \State $d_{\textsf{new}} \gets \textsc{ProvisionDevice}(\mathcal{D}_{\text{empty}}, \mathcal{D}_{\text{base}})$
        \State $d^\star \gets d_{\textsf{new}}$
    \Else
        \State $slack \gets \textsc{ComputeWeightedSlack}(Candidates)$
        \State $d^\star \gets \arg\max_{d\in Candidates} slack$
    \EndIf
    \State $Placement \gets \textsc{AssignOperator}(v,k,d^\star)$
\EndFor
\State \textbf{return} $Placement$
\end{algorithmic}
\end{algorithm}

\Cref{alg:greedy-autoscale} presents the pseudocode for operator provisioning algorithm (\S\ref{sec:design:provisioning}). \Cref{alg:greedy-placement} presents the pseudocode for operator-to-device placement algorithm (\S\ref{sec:design:placement}).

\section{Additional Operator Characterizations}
\label{sec:appendix:characterization}

\S\ref{sec:bg:characterization} presents the characterization study on operator heterogeneity under inference workload changes, and it focuses on compute time, memory usage, and SM core allocation sensitivities. This section analyzes additional types of sensitivities.

\myparagraph{Queueing Characteristics}
Building on the per-operator compute sensitivities, we analyze how operators respond to increasing request load (RPS) using M/M/c queueing theory.
Each operator is modeled as a multi-replica queueing system, where the service rate is $\mu = 1/(\text{op\_latency} \cdot \text{num\_layers})$, derived from the measured GPU execution times, and the arrival rate is $\lambda=$ requests\_per\_second/batch\_size.
Using the Erlang-C formula, we estimate waiting times and determine the minimum number of replicas required to maintain system stability under varying RPS.

\begin{figure}[!t]
    \centering
    \includegraphics[width=\linewidth]{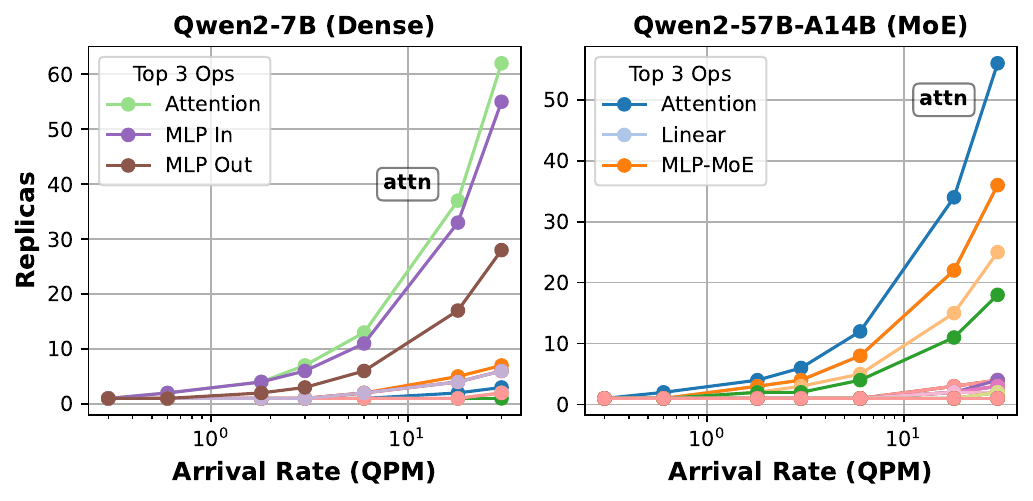}
    \caption{Queueing sensitivity to arrival rate.
    }
    \label{fig:queueing-sensitivity}
\end{figure}

\Cref{fig:queueing-sensitivity} shows heterogeneous queueing sensitivities across operators that closely reflect their compute characteristics.
For attention operators, especially at longer sequence lengths, the number of replicas required grows sharply with increasing RPS, reflecting their high per-token computational cost.
In MoE models like Mixtral, the FusedMoE linear operator dominates compute at short sequence lengths, leading to a pronounced scaling of replicas even for relatively small sequences.
In contrast, lighter operators, such as layer norms and embeddings, exhibit moderate sensitivity to RPS.
Queueing delays increase non-linearly when replication is insufficient, so small reductions in replicas can cause disproportionately high waiting times.
These observations highlight that combining compute profiling and queueing modeling enables precise, operator-specific replication strategies by selectively replicating high-demanding operators (\eg{}, attention) while avoiding overprovisioning lightweight ones, thereby meeting end-to-end latency and throughput targets efficiently.

\boxtakeaway{
Operators exhibit diverse queueing sensitivity with increasing load. Insufficient replication causes non-linear queueing delays, emphasizing the need for operator-specific replication strategies.}
\label{insight:queueing}

\begin{figure}[!t]
    \centering
    \includegraphics[width=\linewidth]{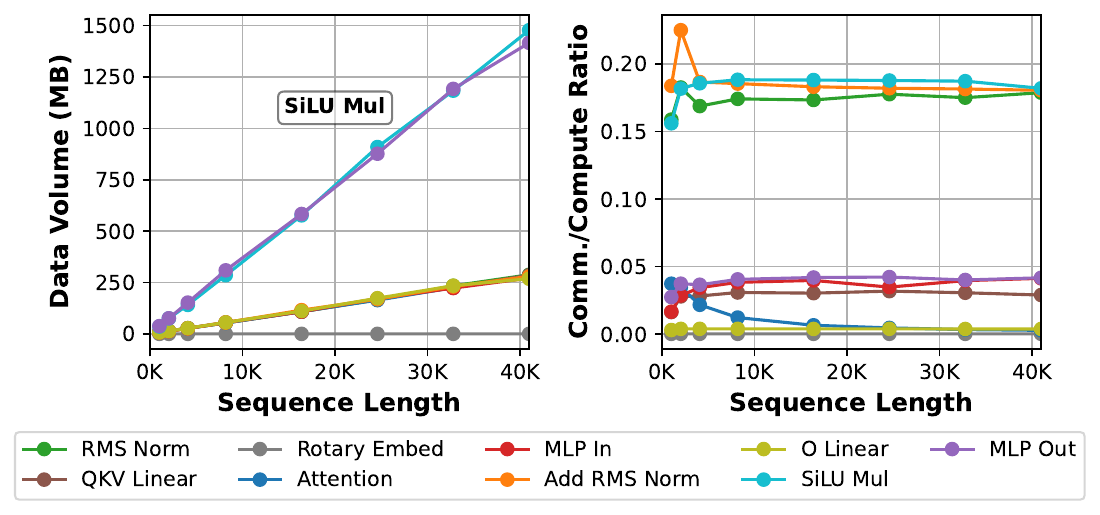}
    \caption{Operator input data volume for Qwen2-7B.}
    \label{fig:io-sensitivity}
\end{figure}

\myparagraph{Dataflow Characteristics}
We analyze data flows by quantifying the communication payload between adjacent operators in the model graph.
Specifically, we perform transient memory profiling to capture how each operator's input-output scale with sequence length and batch size, revealing {that most operators} exhibit linear growth in data volume (as shown in \Cref{fig:io-sensitivity}).
Communication overhead is estimated from transfer latency, which scales proportionally with payload size.
While many operators (\eg{}, element-wise functions) have near-constant per-request data volume, others (\eg{}, attention and linear operators) scale with sequence length.
Comparative analysis of compute versus NVLink transfer time shows transfer overhead reaching $\sim$20\% for certain operators (\eg{}, \texttt{SiLu Mul}) but remaining below 5\% for most operators.

\boxtakeaway{Data volume scales linearly or remains flat with sequence length across operators. Transfer overhead can reach~20\% of compute time, making transfer costs non-trivial when placing operators across devices.}
\label{insight:io}

\section{Model Details}

\begin{table}[t]
\centering
\caption{Models in characterization study and evaluations.}
\label{tab:models}
\resizebox{\linewidth}{!}{%
\begin{tabular}{lccc}
\toprule
\textbf{Model} & \textbf{Model Size} & \textbf{Modality} & \textbf{Architecture} \\
\midrule
Qwen2-7B~\cite{qwen2-7b}        & 7B   & Text & Dense LLM \\
Qwen2-MoE~\cite{qwenmoe}     & 57B (14B)  & Text & MoE LLM \\
Llama3-8B~\cite{llama3}         & 8B   & Text & Dense LLM \\
Mixtral-8$\times$7B~\cite{mixtral} & 47B (13B) & Text & MoE LLM \\
Qwen2.5-VL-32B~\cite{qwenvl}        & 32B  & Visual & Encoder+LLM \\
\bottomrule
\end{tabular}}
\end{table}

Table~\ref{tab:models} details the models used in our characterization study and evaluations. We consider models of two dominant architectures: dense LLMs and mixture-of-experts (MoE) models.

\section{Prefill vs. Decode}
\label{sec:appendix:prefill_vs_decode}

\begin{figure}[!t]
    \centering
    \includegraphics[width=\linewidth]{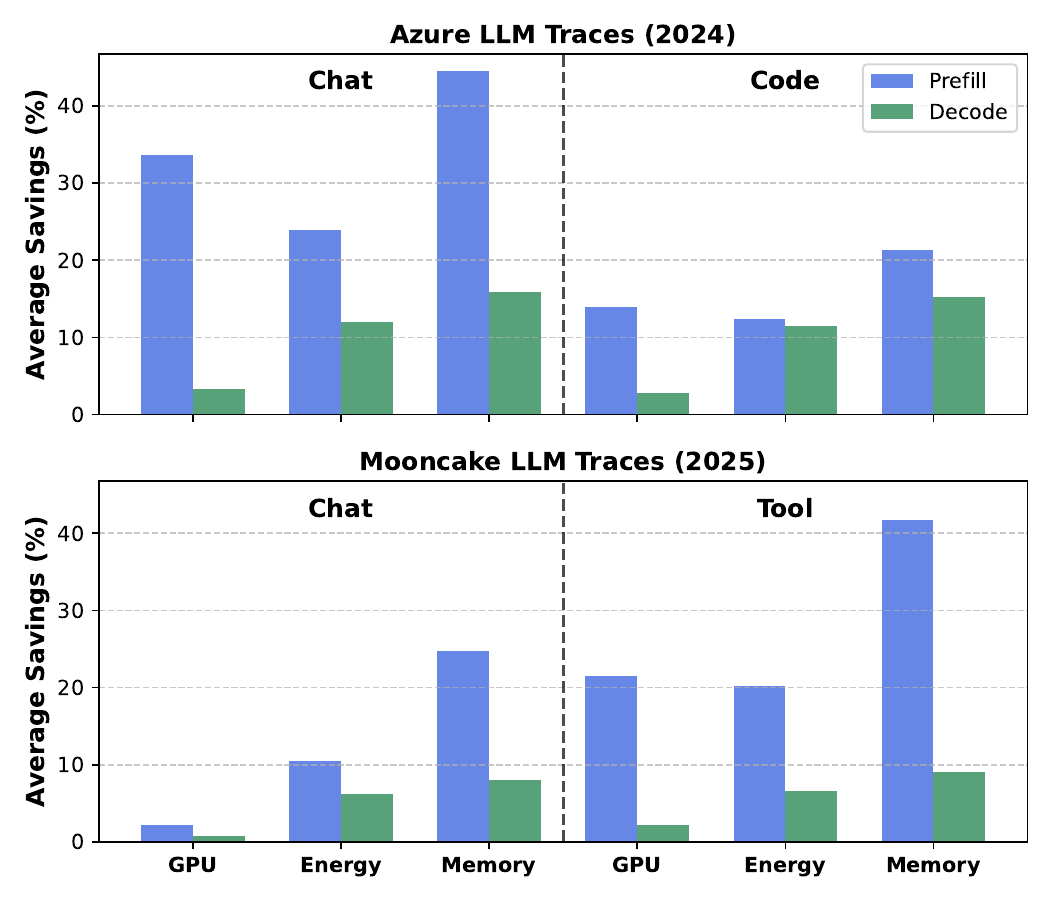}
    \caption{Benefits of operator-level resource management in prefill vs. decode stages for Qwen2-7B.}
    \label{fig:analysis-stage}
\end{figure}

We compare op-level and model-level provisioning behavior across the prefill (known to be compute-bound~\cite{patel2024splitwise}) and decode stages (memory-bound), illustrating how their distinct computational and temporal characteristics drive different scaling needs.
We analyze two production-scale LLM inference traces from (1) Azure LLM inference cluster~\cite{stojkovic2025dynamollm} and (2) Mooncake LLM serving platform~\cite{qin2024mooncake}.
As shown in \Cref{fig:analysis-stage}, op-level resource management yields consistently higher savings during the prefill stage across all workloads.
On Azure traces, prefill achieves up to 35\% GPU, 25\% energy, and 45\% memory savings for chat services, while decode savings remain modest (below 15\%).
Savings in coding services are lower due to their low QPS in the traces.
Similarly, in Mooncake traces, prefill achieves up to 22\% GPU savings, 20\% energy, and 41\% memory, higher than the savings achieved during decode.
These results highlight that prefill stages benefit more from op-level provisioning and scaling due to their denser compute utilization and shorter execution bursts.

\boxinsight{Prefill stages offer substantially higher optimization potential than decode---up to 2--3\texttimes{} greater resource savings, as they are more compute-intensive and bursty, making them ideal targets for fine-grained op-level model provisioning and scaling.}
\label{insight:prefill_vs_decode}

\section{Additional Evaluation Results}
\label{sec:appendix:more-eval}

This section provides detailed evaluation results for the Mixture-of-Experts (MoE) model, Qwen2-57B-A14B. Due to the high operator heterogeneity in MoE architectures (\S\ref{sec:bg:characterization}), monolithic scaling often results in significant resource over-provisioning. Our results demonstrate that \sysname{}'s op-level elasticity is particularly effective for such models, with higher gains compared to serving dense LLMs.

\begin{figure}[t!]
    \centering
    \includegraphics[width=\linewidth]{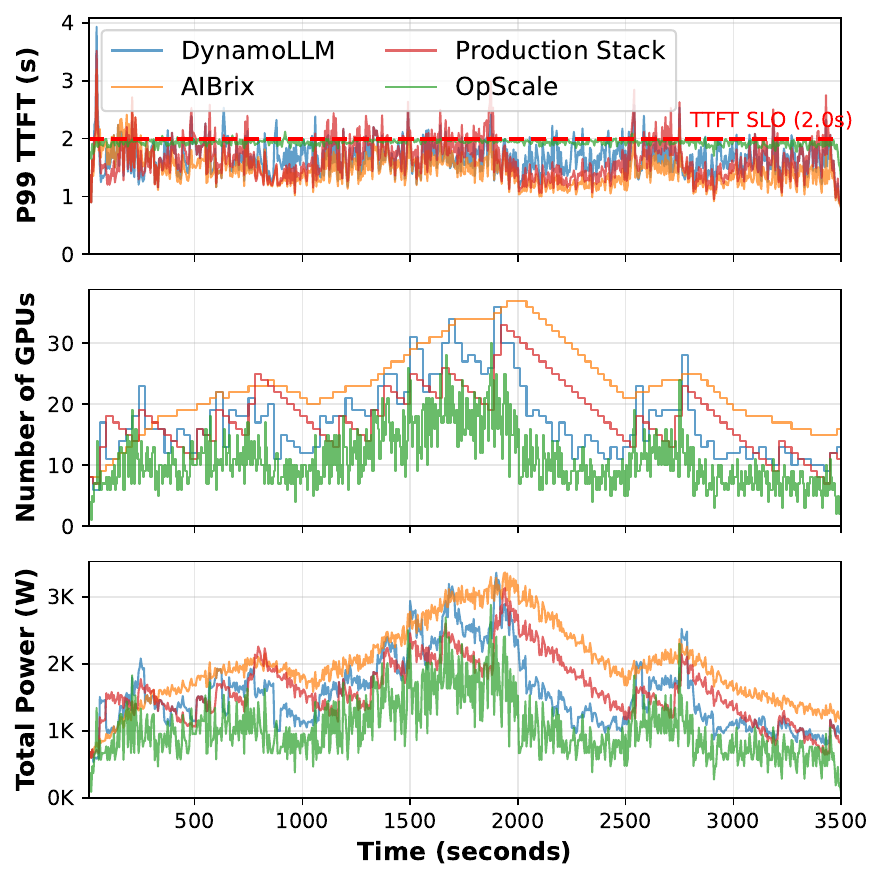}
    \caption{Comparison of GPU usage, request performance (P99 TTFT), and total power usage during autoscaling for Qwen2-57B-A14B (MoE).}
    \label{fig:autoscaling:moe}
\end{figure}

\myparagraph{MoE Autoscaling Performance}
\Cref{fig:autoscaling:moe} illustrates the real-time autoscaling behavior over a 1-hour trace. While all systems aim to keep the P99 TTFT below the 2.0s SLO, the resource footprints vary significantly.
\sysname{} maintains a much tighter envelope around the actual workload demand. As shown in the middle row of \Cref{fig:autoscaling:moe}, \sysname{} (green line) operates with a significantly lower number of GPUs throughout the trace, peaking at roughly 25 GPUs during high-load periods, whereas AIBrix and DynamoLLM frequently exceed 30--35 GPUs to meet the same latency targets.

Crucially, even with fewer resources, \sysname{} maintains a more stable P99 TTFT. While the Production Stack and AIBrix show frequent spikes that touch or exceed the 2.0s SLO line, \sysname{} remains consistently below the threshold. This efficiency is attributed to the system’s ability to scale only the bottlenecked operators rather than replicating the entire 57B parameter model.

\begin{figure}[t!]
    \centering
    \begin{minipage}[t]{0.55\linewidth}
        \centering
        \includegraphics[width=\linewidth]{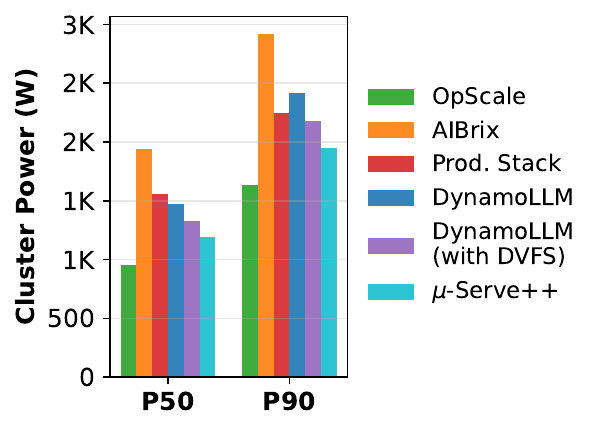}
        \caption{Cluster-level power consumption comparison.}
        \label{fig:barplot_power_moe}
    \end{minipage}
    \hfill
    \begin{minipage}[t]{0.42\linewidth}
        \centering
        \includegraphics[width=\linewidth]{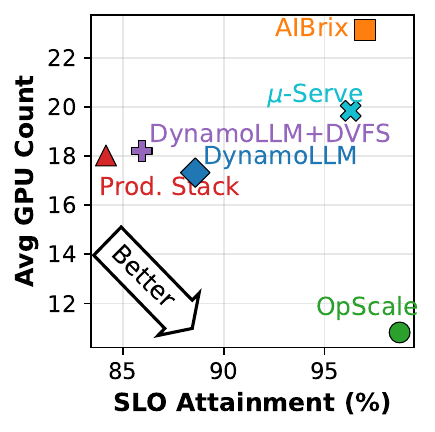}
        \caption{Cost vs. SLO attainment trade-off.}
        \label{fig:scatter_cost_vs_slo_moe}
    \end{minipage}
\end{figure}

\myparagraph{Power and Cost Trade-offs for MoE}
The power consumption benefits observed in the dense model are further amplified in the MoE setting. \Cref{fig:barplot_power_moe} shows the cluster-level power consumption at P50 and P90 levels. \sysname{} with DVFS achieves the lowest power profile, outperforming $\mu$-Serve++ by approximately 15\% at the P90 mark.
Notably, even without DVFS, the base \sysname{} implementation consumes less power than the Production Stack and AIBrix, confirming that reducing the active GPU count through fine-grained scaling is more impactful than model-level power management.

Finally, \Cref{fig:scatter_cost_vs_slo_moe} summarizes the cost-performance Pareto frontier.
\sysname{} occupies the bottom-right corner of the plot, representing the ideal high-SLO-attainment, low-GPU-count quadrant.
Compared to AIBrix, \sysname{} reduces the average GPU count from $\sim$23 to $\sim$11 while simultaneously improving SLO attainment from $\sim$97\% to over 98\%.
While $\mu$-Serve++ and DynamoLLM+DVFS improve upon their respective vanilla baselines, they still require 18--20 GPUs on average. This highlights that for MoE models, the overhead of monolithic replicas is a primary bottleneck that cannot be fully mitigated by frequency scaling or partitioning alone.
These results confirm that operator-level elasticity allows \sysname{} to bypass the all-or-nothing scaling penalty inherent in MoE models, delivering superior performance with a substantially smaller hardware footprint.

\end{document}